\documentclass{article}
\usepackage{latexsym, amsmath, amssymb, mathrsfs, graphicx}
\title{A staggered six-vertex model with non-compact continuum limit}

\author{Yacine Ikhlef${}^{1,2}$, Jesper Jacobsen${}^{1,2}$ and Hubert Saleur${}^{2,3}$ \\[2.0mm]
${}^1$ LPTMS, Universit\'e Paris-Sud, B\^atiment 100, \\
Orsay, 91405, France \\
${}^2$ Service de Physique Th\'eorique, CEA Saclay, \\
Gif Sur Yvette, 91191, France \\
${}^3$ Department of Physics and Astronomy,
University of Southern California, \\
Los Angeles, CA 90089, USA}

\begin{document}

% \myid : 1I : symbol one
\def\myid{{\mathchoice {\rm 1\mskip-4mu l} {\rm 1\mskip-4mu l}
{\rm 1\mskip-4.5mu l} {\rm 1\mskip-5mu l}}}

\newcommand \nearest {\langle ij \rangle}

\maketitle

\begin{abstract}
The antiferromagnetic critical point of the Potts model on the square
lattice was identified by Baxter~\cite{baxter82} as a staggered
integrable six-vertex model. In this work, we investigate 
the integrable structure of this model. It enables us to derive some
new properties, such as the Hamiltonian limit of the model, an
equivalent vertex model, and the structure resulting from the
$Z_2$ symmetry. Using this material, we discuss the low-energy
spectrum, and relate it to geometrical excitations. We also 
compute the critical exponents by solving the Bethe equations for 
a large lattice width $N$. The results confirm that the low-energy 
spectrum is a collection of continua with typical exponent gaps of 
order $(\log N)^{-2}$.
\end{abstract}

\section{Introduction}

Our understanding of  $1+1$ conformal field theories with non compact 
target spaces has improved a great deal in the last few years, thanks 
to the use of geometrical methods \cite{Schomer}, and ideas from
string theory \cite{Ooguri}. The topic is 
of the highest interest in the context of the AdS/CFT duality.

Theories with non compact target spaces also play an important role in 
statistical mechanics. A sophisticated example of this role occurs in 
the supersymmetric approach to phase transitions for non interacting 
disordered electronic systems, where the universality class of the 
transition between plateaux of the integer quantum Hall effect is 
related with the IR limit of a non compact $1+1$  supersigma model at 
$\theta=\pi$ \cite{Zirnbauer}. A more basic  example is 
provided by  Brownian motion and subtle properties thereof, such as the 
(non) intersection exponents \cite{Duplantier}. In both cases, 
the non compacity of the target 
space occurs  because the electron trajectories or 
the random path can visit a given site (edge) an 
infinite number of times. This is in sharp contrast with self 
avoiding models for which almost everything is by now understood, and 
related with ordinary CFTs (essentially, a twisted free boson). 

An obvious strategy to tackle the physics of models with non compact 
target spaces is to start with a lattice model having an infinity of 
degrees of freedom per site/link. For instance, it is easy to 
generalize the usual XXX chain to a non compact representation of 
$SL(2,R)$, and try to use the standard tools of Bethe ansatz, Baxter 
Q-operator, etc, to obtain properties such as gaplessness and critical 
exponents. Despite some serious progress in this direction
\cite{Korchemsky}, the 
problem is far from being closed. 

Another strategy is based on the observation that a non compact 
continuum limit may well arise from a lattice model with finite number of 
degrees of freedom per site/link if the non unitarity is strong 
enough. The two families of examples exhibited so far involve models 
with supergroup symmetries---either models with $OSP(m|2n)$ 
symmetries (such as intersecting loop models) \cite{JRS}
which, in their Goldstone phases can be described in the IR by a 
collection of free bosons and symplectic fermions, or the $SU(1|2)$ 
integrable spin chain with alternating $3$ and $\bar{3}$ 
representations, which was found to be described by the $SU(2|1)$ WZW 
model at $k=1$ \cite{EFS}. In both cases, a  {\sl continuous 
spectrum of critical exponents} 
is found, and the target space does exhibit some non 
compact directions indeed. 

The examples of Brownian motion and self intersecting dense curves 
should convince the reader that non compact target spaces are more 
common and useful than might have been surmised a few years ago. In a 
recent paper, we found \cite{jacobsen-saleur06} that the antiferromagnetic Potts 
model on the square lattice for $Q$ continuous has critical properties 
seemingly involving a twisted non compact boson. This conclusion was based on 
some numerical and analytical evidence, and implied that the well 
known six-vertex model itself might exhibit such an exotic continuum 
limit if properly staggered. The purpose of this paper is to discuss 
these results further, and put them on considerably firmer grounds. 

Indeed, the evidence for a continuous spectrum of critical exponents
is not so easy to obtain from studies of a finite lattice model. What
was really established so far in \cite{JRS,EFS,jacobsen-saleur06} was
that low energy levels appeared with extremely high degeneracies in
the limit of long chains, and that naive calculations of finite size
corrections indicated truly infinite degeneracies. This was---thanks to
complementary arguments, such as mappings onto sigma models, or
abstract construction of WZW theories on supergroups---interpreted as
strong indications for a continuous spectrum in the scaling limit.
Direct evidence was however missing, together with estimates of the
measure of integration on the continuous spectrum, if any. These
issues will be resolved here.

The paper is organized as follows. In section~2, the critical 
antiferromagnetic Potts model and the related staggered 
six-vertex model are defined. Symmetries and limiting cases are studied. 
It is shown in particular that the model coincides with the $OSP(2|2)$ 
model of \cite{JRS} in one limit, and the $SO(4)$ model in another. In 
section~3, the solution of the model using the Bethe ansatz is 
discussed. In section~4, a detailed analysis of the spectrum of 
conformal exponents from the Bethe equations is carried out. Very 
accurate evidence for the existence of a continuous spectrum is 
obtained, together with some information on the measure. This 
information is used in section~5 to relate the results to theoretical 
expectations, in particular those of the supersphere sigma model of 
\cite{JSarboreal}. Elements for a  physical interpretation of the continuous 
spectrum are  proposed  in section~6. Conclusions are gathered in 
section~7.

\section{The staggered six-vertex model and its integrable structure }

\subsection{The general integrable six-vertex model}

% baxter 71 -> eq Bethe, energies libres, etats de Bethe

\begin{figure}
\begin{center}
\includegraphics{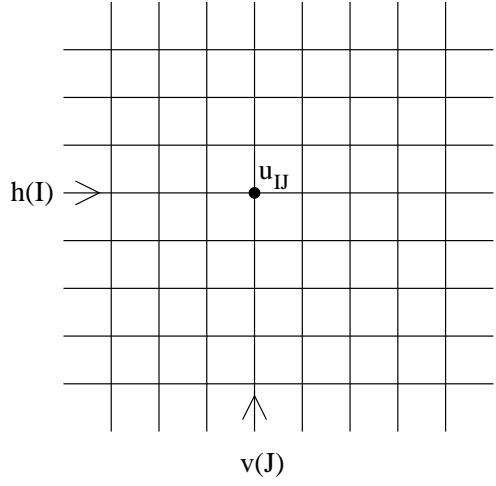}
\caption{The spectral parameter $u_{IJ}$ defined by the line rapidities.}
\label{fig:rapidity}
\end{center}
\end{figure}

\begin{figure}
\begin{center}
\includegraphics[scale=0.8]{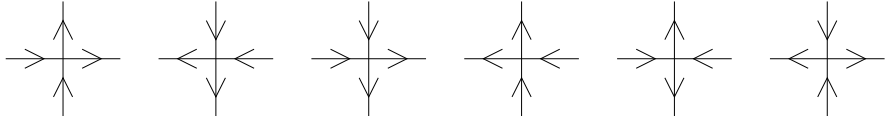}
\caption{The allowed configurations of the six-vertex model.}
\label{fig:six-vertex}
\end{center}
\end{figure}

On the square lattice $\mathcal{L'}$ of $2N$ vertical lines and $M$ 
horizontal lines, associate the complex number $v(J)$ to the $J$-th 
vertical line, and $h(I)$ to the $I$-th horizontal line (see
figure~\ref{fig:rapidity}).
The parameters 
$v(J)$ and $h(I)$ are called \emph{line rapidities}.
On this lattice, define the general inhomogeneous six-vertex model
with local weights given in terms of the difference $u_{IJ}=h(I)-v(J)$.
The weights that satisfy the Yang-Baxter equations are obtained
by taking equations~(1.5)--(1.6) of ref.~\cite{baxter71} and 
performing the substitution~:
\begin{eqnarray}
t, \rho(I), \sigma(J), \kappa(I,J) & \to  &
e^{i \gamma}, e^{i\gamma-2ih(I)}, e^{2iv(J)}, \frac{i}{2} e^{ih(I)-iv(J)-i\gamma} \\
\alpha(I,J), \beta(I,J), \gamma(I,J) & \to &
1, 1, \lambda_{IJ}
\end{eqnarray}
Thus, the weights of the inhomogeneous integrable six-vertex 
model (see figure~\ref{fig:six-vertex}) are~:
\begin{eqnarray}
\omega_1(I,J), \dots,\omega_6(I,J) & = &
\sin \left( \gamma-u_{IJ} \right), \sin \left( \gamma-u_{IJ} \right),
\sin u_{IJ}, \sin u_{IJ}, \notag \\ 
\ & \ & \lambda_{IJ} \ e^{-iu_{IJ}} \sin \gamma,
\left( \lambda_{IJ} \right)^{-1} \ e^{iu_{IJ}} 
\sin \gamma \label{eq:6v-integrable}
\end{eqnarray}
The parameters $\lambda_{IJ}$ must satisfy the additional condition~:
\begin{equation}
\lambda_{I,J+1} \lambda_{I+1,J} = \lambda_{I,J} \lambda_{I+1,J+1}
\end{equation}
The parameters $\lambda_{IJ}$ do not alter the eigenvalues of 
the transfer matrix, thus they play the role of gauge parameters.
The parameter $\Delta$ has the value~:
\begin{equation}
\Delta= - \cos \gamma
\end{equation}

\subsection{The staggered six-vertex model associated 
to the critical antiferromagnetic Potts model}
\label{sec:staggered-6v}

The anisotropic Potts model on the square lattice $\mathcal{L}$ 
is defined by the partition function~:
\begin{equation}  \label{eq:Z-Potts}
Z_{\mathrm{Potts}} = 
\sum_{\{ s_i \}} 
\prod_{\nearest \mathrm{even}} \exp \left[  J_1 \delta(s_i, s_j) \right]
\prod_{\nearest \mathrm{odd}} \exp \left[  J_2 \delta(s_i, s_j) \right]
\end{equation}
where each spin $s_i$ lives on a vertex of $\mathcal{L}$ (white circles), 
and each coupling factor is associated to an edge of $\mathcal{L}$ (dotted
lines). The spin $s_i$ can take $Q$ distinct values.
The sum is over all spin configurations, and each product corresponds
to one type of edge of $\mathcal{L}$.

\begin{figure}
\begin{center}
\includegraphics{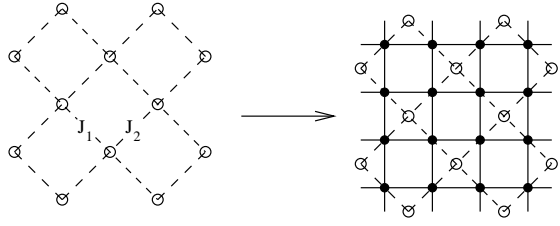}
\caption{The square lattice $\mathcal{L}$ (dotted lines and white circles)
and its medial lattice $\mathcal{L'}$ (full lines and black circles).}
\label{fig:medial}
\end{center}
\end{figure}

When the couplings $J_1$ and $J_2$ are negative, the model is 
antiferromagnetic. In this domain, the critical line separating the
paramagnetic and antiferromagnetic phases is given by the condition~:
\begin{equation} \label{eq:critical1}
\left( e^{J_1} + 1 \right) \left( e^{J_2} + 1 \right) = 4 - Q
\end{equation}
The model defined by eq.~\eqref{eq:Z-Potts} on the square lattice
$\mathcal{L}$ can be mapped to a six-vertex model on the square lattice
$\mathcal{L'}$. This lattice is represented in full lines and black 
circles, and is called the medial lattice of $\mathcal{L}$ 
(see figure~\ref{fig:medial}).
The weights of this six-vertex model depend on the parameters $J_1$, $J_2$
and $Q$. We use the notations~:
\begin{equation}
\sqrt{Q} = 2 \cos \gamma
\end{equation}
\begin{equation}
x_1 = \frac{e^{J_1} - 1}{\sqrt{Q}} \ , \qquad 
x_2 = \frac{e^{J_2} - 1}{\sqrt{Q}}
\end{equation}
The equivalent six-vertex model has weights $\omega_1, \dots , \omega_6$ on
the even vertices and $\omega'_1, \dots , \omega'_6$ on the odd vertices,
where~:
\begin{eqnarray}
\omega_1, \dots , \omega_6 & = &
1,1,x_1,x_1,
e^{i \gamma/2} + x_1 e^{-i \gamma/2}, e^{-i \gamma/2} + x_1 e^{i\gamma/2} 
\label{eq:potts-6v-1} \\
\omega'_1, \dots , \omega'_6 & = &
x_2,x_2,1,1,
e^{-i \gamma/2} + x_2 e^{i \gamma/2}, e^{i \gamma/2} + x_2 e^{-i\gamma/2} 
\label{eq:potts-6v-2} 
\end{eqnarray}
The parameter $\Delta$ is independent of the parameters $x_1$ and $x_2$~:
\begin{equation}
\Delta= 
\frac{\omega_1 \omega_2 + \omega_3 \omega_4 - \omega_5 \omega_6}{2 \omega_1 \omega_3}=
\frac{\omega'_1 \omega'_2 + \omega'_3 \omega'_4 - \omega'_5 \omega'_6}{2 \omega'_1 \omega'_3}=
 - \cos \gamma
\end{equation}
The criticality condition~\eqref{eq:critical1} can be parametrized by~:
\begin{equation} \label{eq:critical2}
x_1 = \frac{\sin u}{\sin(\gamma-u)} \ , \qquad
x_2 = - \frac{\cos(\gamma-u)}{\cos u}
\end{equation}

\begin{figure}
\begin{center}
\includegraphics{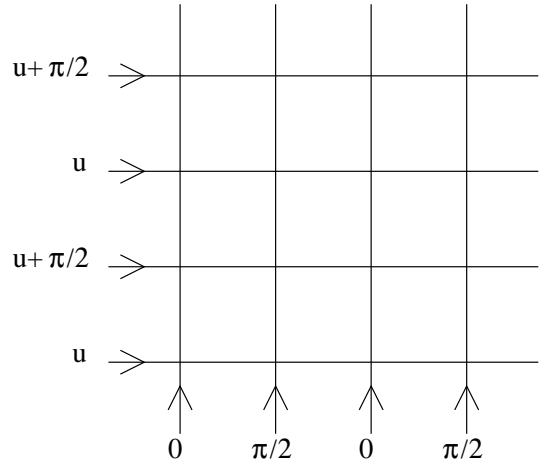}
\caption{The line rapidities corresponding to the antiferromagnetic critical point
of the Potts model}
\label{fig:rapidity-potts-af}
\end{center}
\end{figure}

\begin{figure}
\begin{center}
\includegraphics[scale=0.7]{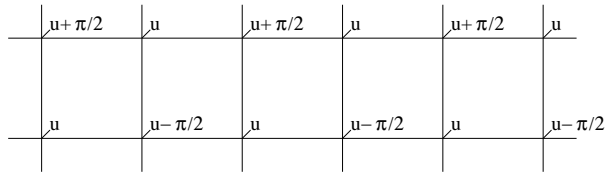}
\caption{The transfer matrix $\mathcal{T}(u)$. The value next to each
vertex is the spectral parameter $u_{IJ}$.}
\label{fig:transfer-matrix}
\end{center}
\end{figure}

When this condition is satisfied,
the weights \eqref{eq:potts-6v-1}--\eqref{eq:potts-6v-2}
of the Potts model correspond to 
a particular case of the integrable six-vertex model
\eqref{eq:6v-integrable}. The rapidity $v(J)$ is equal to 0 (resp. $\pi/2$)
when $J$ is even (resp. odd). The rapidity $h(I)$ is equal to $u$
(resp. $u+\pi/2$) when $I$ is even (resp. odd).
This configuration of the line rapidities is shown in 
figure~\ref{fig:rapidity-potts-af}.
The gauge parameter is set to $\lambda= e^{i \gamma /2}$ at every vertex.
Thus, the partition function of the Potts model at the antiferromagnetic 
critical point is described by the two-row transfer matrix of the 
integrable six-vertex model with the above choice of the rapidities.
We call this matrix $\mathcal{T}(u)$ (see figure~\ref{fig:transfer-matrix}).

\subsection{The $\mathcal{R}$-matrix and the thirty-eight-vertex model}
\label{sec:R-matrix}

\subsubsection{Building block : the $R$-matrix of the six-vertex model}

\begin{figure}
\begin{center}
\includegraphics[scale=1]{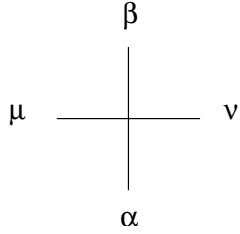}
\caption{The $R$ matrix of the six-vertex model.}
\label{fig:r-matrix}
\end{center}
\end{figure}

\textit{Definition.} The $R$-matrix of the integrable six-vertex 
model is defined
by its matrix element $R^{\nu \beta}_{\alpha \mu}$ equal to
the Boltzmann weight of the configuration shown in 
fig.~\ref{fig:r-matrix}. Let $V$ be the Hilbert
space generated by the vectors~$| \uparrow \rangle, | \downarrow \rangle$.
Then the $R$-matrix is a linear operator mapping the space 
$V_{\mu} \otimes V_{\alpha}$ onto $V_{\beta} \otimes V_{\nu}$.
The matrix elements of the $R$-matrix with spectral parameter
$u_{IJ}=u$ and gauge parameter $\lambda_{IJ}=\lambda$ are the Boltzmann 
weights~\eqref{eq:6v-integrable}. In the basis 
$( | \uparrow \uparrow \rangle, | \uparrow \downarrow \rangle,
   | \downarrow \uparrow \rangle, | \downarrow \downarrow \rangle)$,
the $R$-matrix is~:
\begin{equation} \label{eq:R-matrix-6v}
R(u, \lambda)=
\left(
\begin{array}{llll}
\sin(\gamma-u) & 0 & 0 & 0 \\
0 & \lambda \ e^{-iu} \sin \gamma & \sin u & 0 \\
0 & \sin u & \lambda^{-1} \ e^{iu} \sin \gamma & 0 \\
0 & 0 & 0 & \sin(\gamma-u) 
\end{array}
\right)
\end{equation}
\begin{figure}
\begin{center}
\includegraphics[scale=0.5]{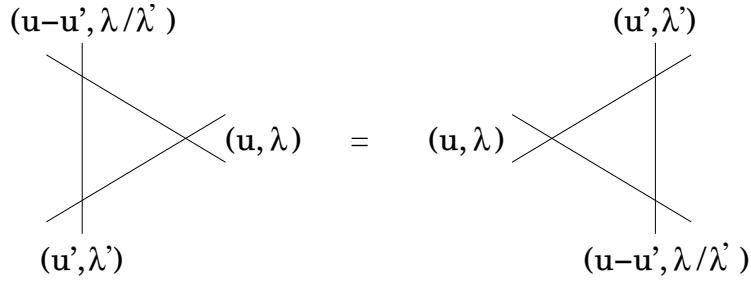}
\caption{The Yang-Baxter relation for the six-vertex R-matrix.}
\label{fig:yang-baxter}
\end{center}
\end{figure}
\\
\\
\textit{Symmetries.} The $R$-matrix satisfies the Yang-Baxter 
equations shown in fig.~\ref{fig:yang-baxter} and the inversion
relation~:
\begin{equation} \label{eq:inversion}
R(u, \lambda)R(-u, \lambda^{-1}) = \sin(\gamma-u) \sin(\gamma+u) \ \myid
\end{equation}
It also has the symmetry property~:
\begin{equation}
R(u+\pi, \lambda) = - R(u, \lambda)
\end{equation}
The $R$-matrix preserves the total magnetization~:
\begin{equation}
\sigma^z_{\mu}+\sigma^z_{\alpha} = \sigma^z_{\beta}+\sigma^z_{\nu} 
\end{equation}
\\
\\
\textit{Relation to the Temperley-Lieb algebra.} When the gauge
is set to $\lambda=1$, the $R$-matrix can be 
written in terms of the Temperley-Lieb generator~$E$~:
\begin{equation} \label{eq:R-E}
R(u, 1) = \sin(\gamma-u) \ \myid + \sin u \ E
\end{equation}
where~:
\begin{equation}
E=
\left(
\begin{array}{llll}
0 & 0 & 0 & 0 \\
0 & e^{-i \gamma} & 1 & 0 \\
0 & 1 & e^{i \gamma} & 0 \\
0 & 0 & 0 & 0
\end{array}
\right)
\end{equation}
In the Hilbert space of row configurations 
$V_1 \otimes \dots \otimes V_{2N}$,
define the operators~:
\begin{equation}
E_m \equiv \myid_1 \otimes \dots \otimes \myid_{m-1} \otimes E 
\otimes \myid_{m+2} \otimes \dots \otimes \myid_{2N} \ , \qquad m=1 \dots 2N
\end{equation}
with a non-trivial action on the space $V_m \otimes V_{m+1}$. This family 
of operators satisfy the Temperley-Lieb algebra~:
\begin{eqnarray}
& \ & (E_m)^2 = \sqrt{Q} \ E_m \label{eq:TL-1}\\ 
& \ & E_m = E_m E_{m+1} E_m = E_m E_{m-1} E_m  \label{eq:TL-2} \\ 
& \ & E_m \ E_{m'} = E_{m'} \ E_m \quad \mathrm{if} \quad |m-m'|>1 \label{eq:TL-3}
\end{eqnarray}
The operators $E_m$ can be expressed in terms of the Pauli matrices~:
\begin{equation} \label{eq:TL-Pauli-1}
E_m = \frac{1}{2} \left[
\sigma_m^x \ \sigma_{m+1}^x + \sigma_m^y \ \sigma_{m+1}^y
- \cos \gamma (\sigma_m^z \ \sigma_{m+1}^z - \myid) 
- i \sin \gamma (\sigma_m^z - \sigma_{m+1}^z ) \right]
\end{equation}
or, in a more compact form~:
\begin{equation} \label{eq:TL-Pauli-2}
E_m = \sigma_m^+ \ \sigma_{m+1}^- + \sigma_m^- \ \sigma_{m+1}^+
+ \frac{1}{2}( \myid - \sigma_m^z \sigma_{m+1}^z ) e^{-i \gamma \sigma_m^z}
\end{equation}

\subsubsection{The $\mathcal{R}$-matrix. Conservation laws.}

The six-vertex model defined in section~\ref{sec:staggered-6v} is
not homogeneous, and the transfer matrix $\mathcal{T}(u)$ is
built using $R$-matrices with different values of $u_{IJ}$.
One can construct a homogeneous model by considering the 
$\mathcal{R}$-matrix, acting on double-edges. 
The double-edges live in the Hilbert space~:
\begin{equation}
\mathcal{V} \equiv V \otimes V
\end{equation}
The $\mathcal{R}$-matrix acts on the product space 
$\mathcal{V} \otimes \mathcal{V}$.

\begin{figure}
\begin{center}
\includegraphics[scale=1]{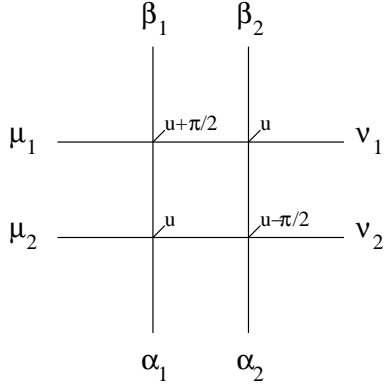}
\caption{The matrix $\mathcal{R}(u)$, defining the antiferromagnetic critical
point of the Potts model.}
\label{fig:r-matrix2}
\end{center} 
\end{figure}

\begin{figure}
\begin{center}
\includegraphics[scale=0.6]{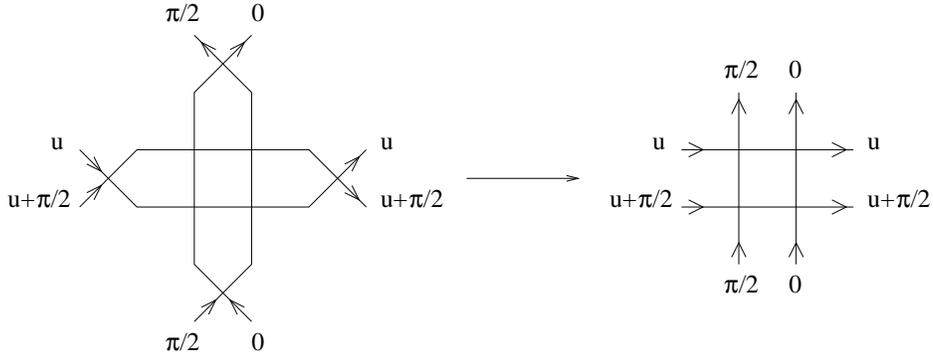}
\caption{A graphical representation of the identity~:
$(c \otimes c)^{-1} \mathcal{R} (c \otimes c) = \mathcal{R}$.}
\label{fig:r-c-commute}
\end{center} 
\end{figure}

As a consequence of the magnetization conservation by $R$, the
$\mathcal{R}$-matrix also preserves the total magnetization~:
\begin{equation}
\sigma^z_{\mu_1} + \sigma^z_{\mu_2} + \sigma^z_{\alpha_1} + \sigma^z_{\alpha_2}
= \sigma^z_{\beta_1} + \sigma^z_{\beta_2} + \sigma^z_{\nu_1} + \sigma^z_{\nu_2} 
\end{equation}
When the gauge parameter is set to $\lambda=1$, another conserved 
quantity can be constructed. Start from the operator~$c$~:
\begin{equation} \label{eq:c}
c \equiv -(\cos \gamma)^{-1} R(\pi/2, 1) = (\cos \gamma)^{-1} R(-\pi/2, 1)
\end{equation}
This operator can be expressed in terms of the Temperley-Lieb generator~:
\begin{equation} 
c = \myid - (\cos \gamma)^{-1} E
\end{equation}
According to the inversion relation~\eqref{eq:inversion},
this operator has the property~:
\begin{equation} \label{eq:c-square}
c^2=\myid
\end{equation}
The $\mathcal{R}$-matrix obeys the conservation rule~:
\begin{equation}\label{eq:r-c-commute}
[\mathcal{R}(u), c \otimes c] = 0
\end{equation}
This is a consequence of the Yang-Baxter equations and the inversion
relation, as shown in fig.~\ref{fig:r-c-commute}.

The eigenvectors of the operator $c$ are~:
\begin{eqnarray}
  \ & | 0 \rangle & = (2 \cos \gamma)^{-1/2} \left( 
    e^{i \gamma /2} | \uparrow \downarrow \rangle 
    -e^{-i \gamma /2} | \downarrow \uparrow \rangle 
  \right) \\
  \ & | \overline{0} \rangle & = (2 \cos \gamma)^{-1/2} \left( 
    e^{-i \gamma /2} | \uparrow \downarrow \rangle 
    +e^{i \gamma /2} | \downarrow \uparrow \rangle 
  \right) \\
  \ & | + \rangle & = | \uparrow \uparrow \rangle \\
  \ & | - \rangle & = | \downarrow \downarrow \rangle
\end{eqnarray}
The eigenspace associated to the eigenvalue $1$ is 
$\{ | + \rangle, | - \rangle, | 0 \rangle \}$,
and the eigen\-space associated to the eigenvalue $-1$ is $\{ | \overline{0} \rangle \}$.

\subsubsection{Mapping to the 38-vertex model}

\begin{figure}
\begin{center}
\includegraphics[scale=0.6]{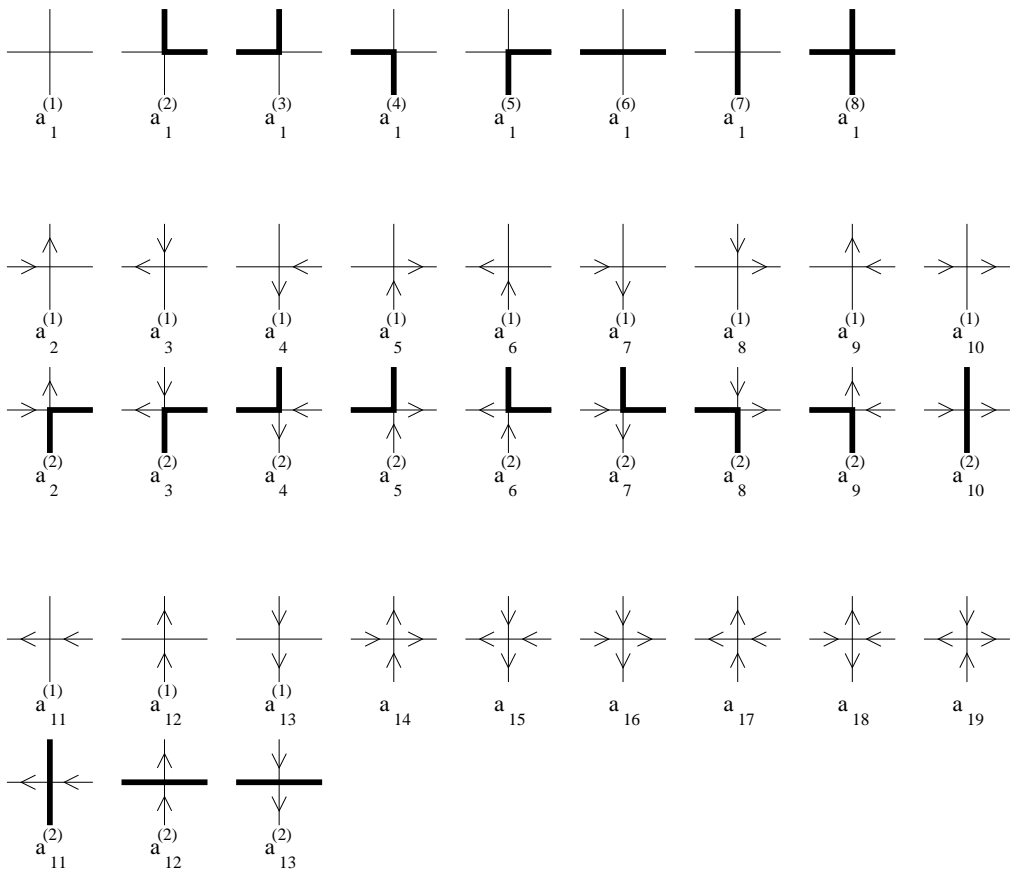}
\caption{The 38-vertex model}
\label{fig:38-vertex}
\end{center} 
\end{figure}

The coefficients of the $\mathcal{R}$-matrix in the basis~:
\begin{equation}
(| + \rangle, | - \rangle, | 0 \rangle, | \overline{0} \rangle) 
\otimes 
(| + \rangle, | - \rangle, | 0 \rangle, | \overline{0} \rangle)
\end{equation}
define the Boltzmann weights of a 38-vertex model on the square
lattice (see figure~\ref{fig:38-vertex}). In this vertex model,
each edge carries an arrow or a thick line. The state $| + \rangle$
(\textit{resp.} $| - \rangle$) is represented by an up or right
(\textit{resp.} down or left) arrow. The state $| 0 \rangle$
is represented by an empty edge, and the state $| \overline{0} \rangle$
by a thick line.
Setting~:
\begin{eqnarray}
\gamma_0 & = & \pi - 2 \gamma \\
u_0 & = & -2 u \\
a_0, b_0, c_0 & = & \sin (\gamma_0-u_0), \sin u_0, \sin \gamma_0
\end{eqnarray}
the weights of the 38-vertex model read~:
\begin{eqnarray}
a_1^{(1)} &=& a_1^{(8)} = -\frac{1}{4} \left[ (c_0)^2 + a_0 b_0 \right] \notag \\
a_1^{(2)} &=& a_1^{(4)} =  \frac{1}{4} b_0 c_0 \notag \\
a_1^{(3)} &=& a_1^{(5)} =  \frac{1}{4} a_0 c_0  \notag \\
a_1^{(6)} &=& a_1^{(7)} = -\frac{1}{4} a_0 b_0 \notag \\
a_2^{(1)} &=& a_4^{(1)} = -\frac{1}{4} e^{-2iu} a_0 c_0 \notag \\
a_2^{(2)} &=& a_4^{(2)} =  \frac{1}{4} a_0 c_0 \notag \\
a_3^{(1)} &=& a_5^{(1)} = -\frac{1}{4} e^{2iu} a_0 c_0 \notag \\
a_3^{(2)} &=& a_5^{(2)} =  \frac{1}{4} a_0 c_0 \notag \\
a_6^{(1)} &=& a_8^{(1)} =  \frac{1}{4} e^{-i (\gamma-2u)} b_0 c_0 \notag \\
a_6^{(2)} &=& a_8^{(2)} =  \frac{1}{4} e^{i \gamma} b_0 c_0 \notag \\
a_7^{(1)} &=& a_9^{(1)} =  \frac{1}{4} e^{i (\gamma-2u)} b_0 c_0 \notag \\
a_7^{(2)} &=& a_9^{(2)} =  \frac{1}{4} e^{-i \gamma} b_0 c_0 \notag \\
a_{10}^{(1)} &=& a_{11}^{(1)} = a_{12}^{(1)} = a_{13}^{(1)} = 
 -\frac{1}{4} a_0 b_0 \notag \\
a_{10}^{(2)} &=& a_{11}^{(2)} = a_{12}^{(2)} = a_{13}^{(2)} = 
 -\frac{1}{4} a_0 b_0 \notag \\
a_{14} &=& a_{15} = -\frac{1}{4} (a_0)^2 \notag \\
a_{16} &=& a_{17} = -\frac{1}{4} (b_0)^2 \notag \\
a_{18} & = & \frac{1}{4} e^{-2iu} c_0 (b_0-a_0) \notag \\
a_{19} & = & \frac{1}{4} e^{2iu} c_0 (b_0-a_0) \notag 
\end{eqnarray}

\subsubsection{The limit $\gamma \to \pi/2$}

If the parameter $\gamma$ is set to $\pi/2$ and the spectral parameter $u$ is fixed,
then the weights of the 38-vertex model are equal to zero, except~:
\begin{eqnarray}
a_1^{(1)} = a_1^{(8)} = a_1^{(6)} = a_1^{(7)} = \frac{1}{4} \sin^2 2u  \notag \\
a_{10}^{(1,2)} = a_{11}^{(1,2)} = a_{12}^{(1,2)} = a_{13}^{(1,2)} = \frac{1}{4} \sin^2 2u  \notag \\
a_{14} = a_{15} = a_{16} = a_{17} = -\frac{1}{4} \sin^2 2u \notag
\end{eqnarray}
The $\mathcal{R}$-matrix is proportional to the ``graded permutation''
$P$~:
\begin{eqnarray}
\mathcal{R} &=& \frac{1}{4} \sin^2 2u \ P 
\qquad (\gamma \to \pi/2, u\ \mathrm{fixed})\\
P | j \rangle | k \rangle &=& (-1)^{g_j g_k} | k \rangle | j \rangle  \ , 
\qquad j,k \in \{ 0,\overline{0},+,- \} \label{eq:perm} \\
g_0,g_{\overline{0}},g_+,g_- &=& 0,0,1,1
\end{eqnarray}
%
%The transfer matrix has the trivial form~:
%\begin{equation}
%\mathcal{T} = 4 \left( \frac{\sin 2u}{2} \right)^{2N} \ P_{\mathrm{even}}
%\end{equation}
%where $P_{\mathrm{even}}$ is the projector on the states with total spin $S$ even.
%
This trivial limit can be avoided by scaling the spectral parameter as~:
\begin{equation}
\gamma = \frac{\pi}{2} + \epsilon \ , \qquad u = \frac{\pi}{2} + \epsilon w
\end{equation}
where $\epsilon \to 0^-$ and $w$ is fixed. In this rescaled limit, the
Boltzmann weights are proportional to $\epsilon^2$.

\begin{figure}
\begin{center}
\includegraphics[scale=0.4]{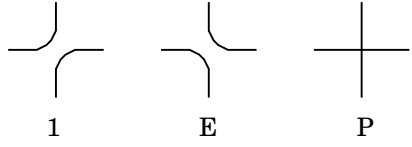}
\caption{The three allowed vertices of the dense intersecting loop
model.}
\label{fig:loop}
\end{center} 
\end{figure}

Note that the states $| 0 \rangle, | \overline{0} \rangle$ become
degenerate, but the following combinations remain non-degenerate~:
\begin{eqnarray}
| 0 \rangle + i | \overline{0} \rangle &=&
e^{i \pi/4} \left[ \tan(-\epsilon/2) \right]^{-1/2} 
(| \uparrow \downarrow \rangle + i | \downarrow \uparrow \rangle) \\
| 0 \rangle - i | \overline{0} \rangle &=&
e^{-i \pi/4} \left[ \tan(-\epsilon/2) \right]^{1/2} 
(| \uparrow \downarrow \rangle - i | \downarrow \uparrow \rangle)
\end{eqnarray}
Denote $\tilde{a}_i^{(j)}$ the matrix elements of 
$\tilde{\mathcal{R}} \equiv \mathcal{R} / (-\epsilon^2)$
in the basis 
$( | 0 \rangle, | \overline{0} \rangle,
| + \rangle, | - \rangle )$. One gets~:
\begin{eqnarray}
\tilde{a}_1^{(1)} = \tilde{a}_1^{(8)} = (1+w-w^2) \notag \\
\tilde{a}_1^{(2)} = \tilde{a}_1^{(4)} = w \notag \\
\tilde{a}_1^{(3)} = \tilde{a}_1^{(5)} = 
\tilde{a}_2^{(1,2)} = \tilde{a}_3^{(1,2)} = 
\tilde{a}_4^{(1,2)} = \tilde{a}_5^{(1,2)} = (1-w) \notag \\
\tilde{a}_1^{(6)} = \tilde{a}_1^{(7)} = 
\tilde{a}_{10}^{(1,2)} = \tilde{a}_{11}^{(1,2)} = 
\tilde{a}_{12}^{(1,2)} = \tilde{a}_{13}^{(1,2)} = w(1-w) \notag \\
\tilde{a}_6^{(1,2)} = \tilde{a}_8^{(1,2)} = iw \notag \\
\tilde{a}_7^{(1,2)} = \tilde{a}_9^{(1,2)} = -iw \notag \\
\tilde{a}_{14} = \tilde{a}_{15} = (1-w)^2 \notag \\
\tilde{a}_{16} = \tilde{a}_{17} = w^2 \notag \\
\tilde{a}_{18} = \tilde{a}_{19} = (1-2w) \notag 
\end{eqnarray}
These weights can be related to an integrable loop model with
$OSP(2|2)$ symmetry. Indeed, the matrix $\tilde{\mathcal{R}}$
can be expressed as a combination of the identity, the 
permutation operator $P$ defined in equation~\eqref{eq:perm},
and the Temperley-Lieb operator $E$.
The latter is defined in tensor notation as a contraction of the spaces
$\mathcal{V}_{\mu}, \mathcal{V}_{\alpha}$ and $\mathcal{V}_{\beta}, \mathcal{V}_{\nu}$
(see figure~\ref{fig:r-matrix2})~:
\begin{equation}
E_{\alpha \mu}^{\nu \beta} = J_{\nu \beta} (J^\dag)_{\alpha \mu}
\end{equation}
where $J$ is the bilinear form in the basis 
$( | 0 \rangle, | \overline{0} \rangle,
| + \rangle, | - \rangle )$~:
\begin{equation}
J=
\left(
\begin{array}{rrcr}
1 & 0 &  0 & 0  \\
0 & 1 &  0 & 0  \\
0 & 0 &  0 & i  \\
0 & 0 & -i & 0
\end{array}
\right) \ , \qquad
J^\dag J = J J^\dag = \myid
\end{equation}
By construction, the operator obeys the Temperley-Lieb algebraic
relations~\eqref{eq:TL-1}, \eqref{eq:TL-2}, \eqref{eq:TL-3}, with~:
\begin{equation}
E^2 = \mathrm{Tr }(J^* J) \ E = 0
\end{equation}
In the block 
$\{ | 0 \rangle \otimes | 0 \rangle, 
| \overline{0} \rangle \otimes | \overline{0} \rangle,
| + \rangle \otimes | - \rangle, 
| - \rangle \otimes | + \rangle \}$, 
the matrix of the operator $E$ is~:
\begin{equation}
E=\left(
\begin{array}{rrrr}
     1 &  1 & -i &  i   \\
     1 &  1 & -i &  i   \\
    -i & -i & -1 &  1   \\
     i &  i &  1 & -1
   \end{array}
\right)
\end{equation}
The matrix $\tilde{\mathcal{R}}$ is equal to~:
\begin{equation} \label{r-osp-2-2}
\tilde{\mathcal{R}} = (1-w) \myid + w E + w(1-w) P
\end{equation}
These weights define an integrable loop model \cite{nienhuis98}, with
three allowed vertices, and a loop fugacity $q=0$. 
The graphical correspondence is shown in figure~\ref{fig:loop}.
Note that according to equation~\eqref{eq:critical2}, the Potts
model is isotropic when $u=\gamma/2 + \pi/4$, which is equivalent to $w=1/2$.
This is also the isotropic point of the loop model~\eqref{r-osp-2-2}.

We conclude that, in the limit $\gamma \to \pi/2$, the staggered
six-vertex model coincides with the $OSP(2|2)$ integrable model. 
In this limit, the arrows represent the fermionic coordinates, and 
the thin and thick lines the bosonic ones. The equivalence in the 
boundary conditions has to be treated with some 
care however: the periodic vertex model corresponds to the 
antiperiodic sector of the $OSP$ system, with an effective 
central charge $c_{\mathrm{eff}} = 1 + (-2) - 24 \times (-1/8)=2$. 
This means that the $OSP$ symmetry is broken in the 
periodic vertex model, where the fermions are twisted - 
\textit{i.e.} they become a complex (Dirac) fermion instead of symplectic 
fermions.

\subsubsection{The limit $\gamma \to 0$}

The approach for this limit is similar to the case $\gamma \to \pi/2$.
Consider the following limit~:
\begin{equation}
\gamma = \epsilon \ , \qquad u = -w \epsilon
\end{equation}
where $\epsilon \to 0^+$ and $w$ is fixed.
It is convenient to perform the change of basis~:
\begin{eqnarray}
 \ & | \tilde{0} \rangle & = i | 0 \rangle \\
 \ & | \tilde{\overline{0}} \rangle & = i | \overline{0} \rangle \\
  \ & | \tilde{+} \rangle & = | + \rangle \\
  \ & | \tilde{-} \rangle & = | - \rangle
\end{eqnarray}
Define a transformation of the $\mathcal{R}$-matrix that does not affect the
partition function : the Boltzmann weights of the vertices are
multiplied by $(-1)$ for each $\pi/2$-turn of the thick lines 
(the weight $a_1^{(8)}$ is not affected). Since the number of such
turns is even for every thick polygon on the lattice, the partition
function is invariant under this transformation.

Denote $\tilde{a}_i^{(j)}$ the matrix elements of
the (rescaled) resulting matrix
$\tilde{\mathcal{R}} \equiv \mathcal{R} / (-\epsilon^2)$
in the basis 
$( | \tilde{0} \rangle, | \tilde{\overline{0}} \rangle,
| \tilde{+} \rangle, | \tilde{-} \rangle )$. One gets~:
\begin{eqnarray}
\tilde{a}_1^{(1)} = \tilde{a}_1^{(8)} = (1+w+w^2) \notag \\
\tilde{a}_1^{(2)} = \tilde{a}_1^{(4)} = -w \notag \\
\tilde{a}_1^{(3)} = \tilde{a}_1^{(5)} = 
\tilde{a}_2^{(1,2)} = \tilde{a}_3^{(1,2)} = 
\tilde{a}_4^{(1,2)} = \tilde{a}_5^{(1,2)} = (1+w) \notag \\
\tilde{a}_1^{(6)} = \tilde{a}_1^{(7)} = 
\tilde{a}_{10}^{(1,2)} = \tilde{a}_{11}^{(1,2)} = 
\tilde{a}_{12}^{(1,2)} = \tilde{a}_{13}^{(1,2)} = w(1+w) \notag \\
\tilde{a}_6^{(1,2)} = \tilde{a}_8^{(1,2)} = -w \notag \\
\tilde{a}_7^{(1,2)} = \tilde{a}_9^{(1,2)} =- w \notag \\
\tilde{a}_{14} = \tilde{a}_{15} = (1+w)^2 \notag \\
\tilde{a}_{16} = \tilde{a}_{17} = w^2 \notag \\
\tilde{a}_{18} = \tilde{a}_{19} = 1 \notag 
\end{eqnarray}
The corresponding loop model is constructed with two generators $P$
and $E$.
The permutation operator $P$ is defined as~:
\begin{equation}
P_{\beta \nu \mu \alpha} = \delta_{\beta \alpha} \delta_{\nu \mu}
\end{equation}
The Temperley-Lieb operator $E$ with parameter $\sqrt{Q}=4$
is built using the simple contraction~:
\begin{equation}
E_{\beta \nu \mu \alpha} = J_{\nu \beta} J_{\alpha \mu}
\end{equation}
where~:
\begin{equation}
J=
\left(
\begin{array}{rrcr}
1 & 0 & 0 & 0  \\
0 & 1 & 0 & 0  \\
0 & 0 & 0 & 1  \\
0 & 0 & 1 & 0
\end{array}
\right) \ , \qquad
J^2 = \myid
\end{equation}
The matrix $\tilde{\mathcal{R}}$ is equal to~:
\begin{equation} \label{r-so-4}
\tilde{\mathcal{R}} = (1+w) \myid - w E + w(1+w) P
\end{equation}
These are the integrable weights of the loop model defined in
\cite{nienhuis98} with a loop fugacity $q=4$. 
Note that, in this regime, the Potts model is anisotropic for
any value of $w$. The loop model itself is isotropic when $w=-1/2$~:
the relative weight for loop crossings is then equal to 
$-1/2$. Following \cite{nienhuis98}, this model is a graphical 
version of the $SO(4)$ integrable model based on the vector 
representation. Using that $so(4)=sl(2)+sl(2)$ one can expect it to 
decouple into two copies of the isotropic six-vertex model or XXX spin 
chain, a feature we will confirm when discussing the associated 
hamiltonian or Bethe equations.

\subsection{Transfer matrices}

\begin{figure}
\begin{center}
\includegraphics[scale=0.4]{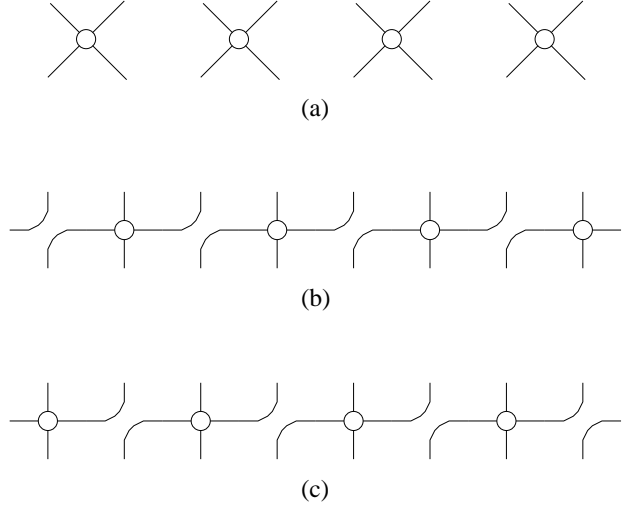}
\caption{Diagrams defining the transfer matrices $C$ (a), $e^{-iP} C$ (b)
and $C e^{-iP}$ (c), for $N=4$. The white circles represent $c$ operators.}
\label{fig:transfer-matrix-one-row}
\end{center} 
\end{figure}

In this section, we set the gauge to $\lambda=1$, so that
the transfer matrices can be related to the generators of the
Temperley-Lieb algebra.
\\
\\
\textit{One-row transfer matrix.} The two-row transfer matrix 
$\mathcal{T}(u)$ is the product of two
one-row transfer matrices~: $\mathcal{T}(u) = T(u+\pi/2) \ T(u)$.
(see fig.~\ref{fig:transfer-matrix}).
As a consequence of the Yang-Baxter equations 
and the inversion relation, when periodic boundar
conditions are imposed in the
horizontal direction, the one-row transfer matrices
commute~:
\begin{equation} \label{eq:t-t-commute}
[T(u), T(u')]=0
\end{equation}
The one-row transfer matrices have the property~:
\begin{equation} \label{eq:t1-t2}
T(u+\pi/2) = (-1)^N e^{iP} T(u) e^{-iP}
\end{equation}
Define the operator $e^{-iP}$ as the shift
of all vertical edges by one site to the right. 
The operator $c$ defined by equation~\eqref{eq:c} is used to build
a conserved quantity of the transfer matrix. Denoting 
$c_m$ the charge operator
acting on the space $V_m \otimes V_{m+1}$, the global charge is~:
\begin{equation} \label{eq:C}
C \equiv c_1 \times  c_3 \times \dots \times c_{2N-1}
\end{equation}
\begin{equation}
C^2= \myid
\end{equation}
If $u=0$, the transfer matrices $T(u), T(u+\pi/2)$ become~:
\begin{equation} \label{eq:t1}
T(0) = \left( \frac{1}{2} \sin 2 \gamma \right)^N \ e^{-iP} \ C
\end{equation}
\begin{equation} \label{eq:t2}
T(\pi/2) = \left(-\frac{1}{2} \sin 2 \gamma \right)^N \ C \ e^{-iP}
\end{equation}
The matrices~\eqref{eq:C}, \eqref{eq:t1}, \eqref{eq:t2} are represented
in figure~\ref{fig:transfer-matrix-one-row}.
\\
\\
\textit{Conservation laws for $\mathcal{T}$.} As a consequence 
of the conservation of the total magnetization and the ``charge''
$c$ by the $\mathcal{R}$-matrix, the transfer matrix $\mathcal{T}$
preserves the total magnetization~:
\begin{equation}
S=\frac{1}{2} (\sigma^z_1 + \dots + \sigma^z_{2N})
\end{equation}
and the ``charge'' $C$ defined above. Since $\mathcal{T}(u)$ also
commutes with the transfer matrix $T(0)$, $\mathcal{T}(u)$ is
invariant under the action of the one-site shift operator $e^{-iP}$.

\subsection{Hamiltonian limit of the transfer matrix}

\begin{figure}
\begin{center}
\includegraphics[scale=0.4]{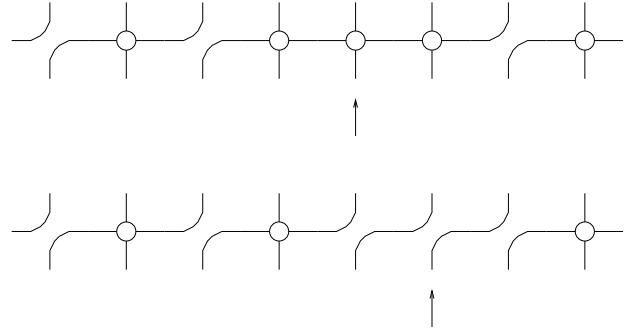}
\caption{Diagrams defining $t_{2j-1}$ and $t_{2j}$ of 
eq.~\eqref{eq:delta-t}, for $N=4$ and $j=3$. The white circles 
represent $c$ operators. The arrow points to the vertex that differs
from $T(0)$.}
\label{fig:delta-t}
\end{center} 
\end{figure}

To compute the derivative of the matrix $T(u)$ at the points 
$u=0, u=\pi/2$, it is convenient to write this matrix in terms
of the $R$-matrix~:
\begin{equation}
  T^{\beta_1 \dots \beta_{2N}}_{\alpha_1 \dots \alpha_{2N}}(u) =
  \sum_{\mu_1, \dots, \mu_{2N}}
  R^{\mu_2 \beta_1}_{\alpha_1 \mu_1}(u) 
  R^{\mu_3 \beta_2}_{\alpha_2 \mu_2}(u-\pi/2)
  \dots
  R^{\mu_{2N} \beta_{2N-1}}_{\alpha_{2N-1} \mu_{2N-1}}(u) 
  R^{\mu_1 \beta_{2N}}_{\alpha_{2N} \mu_{2N}}(u-\pi/2) \notag
\end{equation}
Denote by $\delta$ the derivative with respect to $u$.
According to eq.~\eqref{eq:R-E}~:
\begin{equation}
\delta R(u) = -\cos(\gamma-u) \myid + \cos u \ E
\end{equation}
Differentiating the previous expression yields~:
\begin{equation} \label{eq:delta-t}
\delta T(0) = \left( \frac{1}{2} \sin 2 \gamma \right) ^{N-1}
\sum_{j=1}^{N} ( -\cos^2 \gamma \ t_{2j-1} + \sin^2 \gamma \ t_{2j} )
\end{equation}
where the matrices $t_{2j-1}$ and $t_{2j}$ are defined by the
diagrams on fig.~\ref{fig:delta-t}.
Using the graphical representation of the matrices $T(0), t_{2j-1}, t_{2j}$
in figures~\ref{fig:transfer-matrix-one-row}--\ref{fig:delta-t} and
the property~\eqref{eq:c-square}, one gets the intermediate results~:
\begin{eqnarray}
\left[ T(0) \right]^{-1} t_{2j-1} &=& \left( \frac{1}{2} \sin 2 \gamma \right)^{-1}
c_{V_{2j-3} \otimes V_{2j-2}} \ c_{V_{2j-2} \otimes V_{2j-1}} \ c_{V_{2j-3} \otimes V_{2j-2}} \\
\left[ T(0) \right]^{-1} t_{2j} &=& \left( \frac{1}{2} \sin 2 \gamma \right)^{-1}
c_{V_{2j-1} \otimes V_{2j}}
\end{eqnarray}
The Hamiltonians $H_1, H_2$ are defined as the logarithmic derivatives
of the transfer matrix $T(u)$ at the points $u=0, \pi/2$.
\begin{eqnarray}
H_1 & \equiv & \frac{1}{2} \sin 2 \gamma \ [T(0)]^{-1} \delta T(0) \notag \\
\ & = & \sum_{j=1}^{N} [ 
 -\cos 2 \gamma \ \myid + \cos \gamma \ (E_{2j-1}+E_{2j})
 - (E_{2j}E_{2j-1}+E_{2j-1}E_{2j}) ] \notag \\
\ & \ & \
\end{eqnarray}
Using eq.~\eqref{eq:t1-t2},
\begin{eqnarray}
H_2 & \equiv & \frac{1}{2} \sin 2 \gamma \ [T(\pi/2)]^{-1} \delta T(\pi/2) \notag \\
\ & = & e^{iP} H_1 e^{-iP} \notag \\
\ & = & \sum_{j=1}^{N} [ 
-\cos 2 \gamma \ \myid + \cos \gamma \ (E_{2j}+E_{2j+1})
 \ - (E_{2j+1}E_{2j}+E_{2j}E_{2j+1}) ] \notag \\
\ & \ & \
\end{eqnarray}
The Hamiltonian associated with the two-row transfer matrix $\mathcal{T}(u)$
is defined as~:
\begin{equation}
\mathcal{H} \equiv \frac{1}{2} \sin 2 \gamma  \ 
[\mathcal{T}(0)]^{-1} \delta \mathcal{T}(0)
\end{equation}
According to the commutation relations~\eqref{eq:t-t-commute},
\begin{eqnarray}
\mathcal{H}  & = & H_1 + H_2  \notag \\
\ & = & \sum_{m=1}^{2N} [
-\cos 2 \gamma \ \myid + 2 \cos \gamma \ E_m - (E_{m+1}E_{m}+E_{m}E_{m+1}) ] \label{eq:h}
\end{eqnarray}
Note that the Hamiltonian $\mathcal{H}$ is the sum of two commuting
parts~:
\begin{equation}
\mathcal{H} = H_1 + H_2 \ , \qquad [H_1, H_2]=0
\end{equation}
A consistent decomposition of the momentum operator is~:
\begin{equation}
e^{-2iP} = (e^{-iP} C) \times (C e^{-iP}) = (C e^{-iP}) \times (e^{-iP} C) 
\end{equation}
Indeed, the operators $H_1, H_2, (e^{-iP} C), (C e^{-iP})$ all commute
with one another.

Using the expression~\eqref{eq:TL-Pauli-2} of the Temperley-Lieb 
generators, the Hamiltonian~\eqref{eq:h} is expressed in terms of 
Pauli matrices~:
\begin{eqnarray}
\mathcal{H} &=& \sum_{m=1}^{2N} \big[ 
-(\sigma_m^+ \sigma_{m+2}^- + \sigma_m^- \sigma_{m+2}^+)
+ \sin^2 \gamma \ \sigma_m^z \sigma_{m+1}^z 
- \frac{1}{2} \sigma_m^z \sigma_{m+2}^z \notag \\
& & + i \sin \gamma \ (\sigma_{m-1}^z - \sigma_{m+2}^z)
(\sigma_m^+ \sigma_{m+1}^- + \sigma_m^- \sigma_{m+1}^+) 
-\frac{1}{2} \cos 2 \gamma \ \myid
\big] \label{eq:H-Pauli}
\end{eqnarray}

In the limit $\gamma \to 0$, the Hamiltonian~\eqref{eq:H-Pauli}
describes, as expected from the $SO(4)$ identification,  two decoupled ferromagnetic XXX spin-chains~:
\begin{equation}
\mathcal{H} = 
-\frac{1}{2} \sum_{j=1}^N 
(\overrightarrow{\sigma}_{2j-1} \cdot \overrightarrow{\sigma}_{2j+1} + \myid)
-\frac{1}{2} \sum_{j=1}^N 
(\overrightarrow{\sigma}_{2j} \cdot \overrightarrow{\sigma}_{2j+2} + \myid)
\qquad (\gamma=0)
\end{equation}

\section{Bethe equations, Bethe states and eigenvalues}

When periodic conditions are imposed in the horizontal direction, the
staggered six-vertex model is solvable by 
Bethe ansatz \cite{baxter71}. We call $r$ the total number of particles.

\subsection{Bethe equations for two types of particles}

The Bethe ansatz equations are~:
\begin{equation} \label{eq:BE-H-1}
\forall j \in \{ 1, \dots, r\} \qquad \exp \left[ 2 i N k(\alpha_j) \right] = 
- \prod_{l=1}^{r} \exp \left[ -i \phi(\alpha_j, \alpha_l) \right]
\end{equation}
The one-particle momentum and the scattering amplitude are given by~:
\begin{equation}
\exp \left[ 2 i k(\alpha) \right] = 
\frac{\sinh (\alpha+i\gamma)}{\sinh (\alpha-i\gamma)}
\end{equation}
\begin{equation}
\exp \left[ i \phi(\alpha, \alpha') \right] = 
\frac{\sinh \frac{1}{2} (\alpha-\alpha'-2i\gamma)}
     {\sinh \frac{1}{2} (\alpha-\alpha'+2i\gamma)}
\end{equation}
The roots $\alpha_j$ describing the ground state and the physical
excitations are expected to sit on the two lines 
$\mathrm{Im}(\alpha) = \pm \pi/2$. Define two types of particles~:
\begin{eqnarray}
\alpha_j^+ & = & \lambda_j + i \pi / 2 , \qquad j=1, \dots r_+ \\
\alpha_j^- & = & \mu_j - i \pi / 2 , \qquad j=1, \dots r_-
\end{eqnarray}
The one-particle momenta are~:
\begin{equation}
\exp \left[ 2i \tilde{k}(\lambda) \right] \equiv 
\exp \left[ 2i k \left(\lambda \pm i \frac{\pi}{2} \right) \right] = 
\frac {\cosh(\lambda+ i \gamma)} {\cosh(\lambda- i \gamma)}
\end{equation}
The scattering amplitude between two particles of the same
type is~:
\begin{equation}
\exp \left[ i \phi_1 ( \lambda , \lambda' ) \right] \equiv
\exp \left[ i \phi \left( \lambda \pm i \frac{\pi}{2}, 
  \lambda' \pm i \frac{\pi}{2} \right) \right] = 
\frac {\sinh \frac{1}{2} (\lambda - \lambda' - 2i \gamma)}
      {\sinh \frac{1}{2} (\lambda - \lambda' + 2i \gamma)}
\end{equation}
The scattering amplitude between two particles of different
types is~:
\begin{equation}
\exp \left[ i \phi_{-1} ( \lambda , \lambda' ) \right] \equiv
\exp \left[ i \phi \left( \lambda \pm i \frac{\pi}{2}, 
  \lambda' \mp i \frac{\pi}{2} \right) \right] = 
\frac {\cosh \frac{1}{2} (\lambda - \lambda' - 2i \gamma)}
      {\cosh \frac{1}{2} (\lambda - \lambda' + 2i \gamma)}
\end{equation}
Define the shifted scattering amplitudes as odd functions of $(\lambda-\lambda')$~:
\begin{equation}
\exp \left[ i \Theta_1(\lambda , \lambda' ) \right] = 
 - \exp \left[ i \phi_1(\lambda , \lambda' ) \right] 
\end{equation}
\begin{equation}
\exp \left[ i \Theta_{-1} (\lambda , \lambda' ) \right] = 
  \exp \left[ i \phi_{-1} (\lambda , \lambda' ) \right] 
\end{equation}
The Bethe equations~\eqref{eq:BE-H-1} split into two sets~:
\begin{equation} \label{eq:BE-H-2}
\forall j \in \{ 1, \dots, r_+\} \quad 
e^{2i N \tilde{k}(\lambda_j)} = 
(-1)^{r_+ - 1} 
\prod_{l=1}^{r_+} e^{-i \Theta_1(\lambda_j , \lambda_l)}
\prod_{l=1}^{r_-} e^{-i \Theta_{-1}(\lambda_j , \mu_l)} 
\end{equation}
\begin{equation} \label{eq:BE-H-3}
\forall j \in \{ 1, \dots, r_-\} \quad 
e^{2i N \tilde{k}(\mu_j)} = 
(-1)^{r_- - 1} 
\prod_{l=1}^{r_+} e^{-i \Theta_{-1}(\mu_j , \lambda_l)}
\prod_{l=1}^{r_-} e^{-i \Theta_1(\mu_j , \mu_l)} 
\end{equation}
Take the logarithm of equations~\eqref{eq:BE-H-2}--\eqref{eq:BE-H-3}~:
\begin{equation} \label{eq:BE-H-4}
\forall j \in \{ 1, \dots, r_+\} \quad 
2N \tilde{k}(\lambda_j) = 2 \pi I_j^+ 
- \sum_{l=1}^{r_+} \Theta_1(\lambda_j , \lambda_l)
- \sum_{l=1}^{r_-} \Theta_{-1}(\lambda_j , \mu_l)
\end{equation}
\begin{equation} \label{eq:BE-H-5}
\forall j \in \{ 1, \dots, r_-\} \quad 
2N \tilde{k}(\mu_j) = 2 \pi I_j^- 
- \sum_{l=1}^{r_+} \Theta_{-1}(\mu_j , \lambda_l)
- \sum_{l=1}^{r_-} \Theta_1(\mu_j , \mu_l)
\end{equation}
The ``Bethe integers'' $I_j^\pm$ follow the following rules~: if $r_+$ 
(\textit{resp.} $r_-$) is even, then all the
$I_j^+$ (\textit{resp.} $I_j^-$) are half-odd integers; if $r_+$ 
(\textit{resp.} $r_-$) is odd, then all the
$I_j^+$ (\textit{resp.} $I_j^-$) are integers.
Summing equations~\eqref{eq:BE-H-4}--\eqref{eq:BE-H-5} over $j$ and
recalling that the functions $\Theta_{\pm 1}$ are odd, one relates
the total momentum to the Bethe integers~:
\begin{equation}
k_{\mathrm{tot}} =
\sum_{j=1}^{r_+} \tilde{k}(\lambda_j) + \sum_{j=1}^{r_-}
\tilde{k}(\mu_j) =
 \frac{\pi}{N} \left( 
\sum_{j=1}^{r_+} I^+_j + \sum_{j=1}^{r_-} I^-_j \right)
\end{equation}
The one-particle momenta and the scattering amplitudes can be written~:
\begin{equation}
\tan \left[ \tilde{k}(\lambda) \right] = \tanh \lambda \ \tan \gamma 
\end{equation}
\begin{equation}
\tan \left[ \frac{\Theta_1(\lambda, \lambda') }{2} \right] = 
\tanh \frac{1}{2} (\lambda - \lambda') \ \mathrm{cotan} \gamma 
\end{equation}
\begin{equation}
\tan \left[ \frac{\Theta_{-1}(\lambda, \lambda') }{2} \right] = 
- \tanh \frac{1}{2} (\lambda - \lambda') \ \tan \gamma 
\end{equation}

\subsection{Bethe states}

\emph{Form of the Bethe states.} The one-particle states 
$\varphi_\alpha$ are ``inhomogeneous plane-waves'', represented in
real space as~:
\begin{eqnarray}
\varphi_\alpha(2m-1) & = & \cosh \frac{1}{2} (\alpha - i \gamma) \ e^{2i k(\alpha) m} \\
\varphi_\alpha(2m) & = & -\sinh \frac{1}{2} (\alpha + i \gamma) \ e^{2i k(\alpha) m}
\end{eqnarray}
with $m=1, \dots, N$ defines the position of the particle. 
The general Bethe state is given by the linear combination~:
\begin{equation}
\varphi_{\alpha_1, \dots, \alpha_r}(x_1, \dots, x_r) = 
\sum_P A_{p_1, \dots, p_r} \varphi_{\alpha_{p_1}}(x_1) \dots \varphi_{\alpha_{p_r}}(x_r)
\end{equation}
where the sum is over all permutations $P=\{ p_1, p_2, \dots, p_r \}$ 
of the integers $1, \dots, r$, and
$x_1, \dots, x_r$ are the positions of the particles.
The coefficients $A_{p_1, \dots, p_r}$ are given in terms of the scattering
amplitudes $s_{j,l}$~:
\begin{equation}
A_{p_1, \dots, p_r} = \epsilon_P \prod_{i < j} s_{p_j,p_i}
\end{equation}
\begin{equation}
s_{j,l} = \sinh \frac{1}{2} (\alpha_j - \alpha_l - 2i \gamma)
\end{equation}
where $\epsilon_P$ is the signature of the permutation $\{p_1, \dots,
p_r\}$. In particular, when the roots $\alpha_j$ sit on the two lines
$\mathrm{Im} (\alpha) = \pm \pi / 2$, we introduce another notation 
for the Bethe state~:
\begin{equation} \label{eq:phi-two-lines}
\varphi_{(\lambda_1, \dots, \lambda_{r_+} | \mu_1, \dots, \mu_{r_-})} 
\equiv 
\varphi_{\lambda_1+i \pi/2, \dots, \lambda_{r_+}+i \pi/2, 
\mu_1-i \pi/2, \dots, \mu_{r_-}-i \pi/2} 
\end{equation}
\\
\\
\emph{Symmetries.} By construction, the states 
$\varphi_{\alpha_1, \dots, \alpha_r}$ are
antisymmetric under the exchange of the $\alpha_j$.
A shift $\alpha_j \to \alpha_j + 2 i \pi$ results in a phase factor~:
\begin{equation}
\varphi_{\alpha_1, \alpha_2, \dots, \alpha_j+2i\pi, \dots, \alpha_r} = 
(-1)^r \ \varphi_{\alpha_1, \alpha_2, \dots, \alpha_j, \dots, \alpha_r}
\end{equation}
\\
\\
\emph{Action of $C$.} The action of the operator $C$ on the Bethe 
states is a shift $\alpha_j \to \alpha_j + i \pi$~:
\begin{equation} \label{eq:action-C-1}
C \varphi_{\alpha_1, \dots, \alpha_r} = 
(-i)^r \ \varphi_{\alpha_1+i\pi, \dots, \alpha_r+i\pi}
\end{equation}
The operator $C$, acting on the states~\eqref{eq:phi-two-lines}, 
exchanges the two lines~:
\begin{equation} \label{eq:action-C-2}
C \varphi_{(\lambda_1, \dots, \lambda_{r_+} | \mu_1, \dots, \mu_{r_-})} = 
e^{\frac{i \pi}{2} (r_+ - r_-)} \ \varphi_{(\mu_1, \dots, \mu_{r_-} | \lambda_1, \dots, \lambda_{r_+})}
\end{equation}
A special class of states (to be discussed in
section~\ref{sec:common}) is defined by the following 
constraint on the Bethe integers~:
\begin{equation}
r_+ = r_- = r
\end{equation}
\begin{equation}
\forall j \in \{ 1, \dots, r \} \ , \qquad I^+_j = I^-_j = I_j
\end{equation}
These states are eigenvectors of the operator $C$, with eigenvalue one~:
\begin{equation} \label{eq:C-phi-sym}
C \varphi_{(\lambda_1, \dots, \lambda_r | \lambda_1, \dots, \lambda_r)}
= \varphi_{(\lambda_1, \dots, \lambda_r | \lambda_1, \dots, \lambda_r)}
\end{equation}
The other eigenvectors of $C$ are given by the linear combinations~:
\begin{equation}
\varphi_{(\lambda_1, \dots, \lambda_{r_+} | \mu_1, \dots, \mu_{r_-})}^{\pm}= 
\frac{1}{\sqrt{2}} \left(
\varphi_{(\lambda_1, \dots, \lambda_{r_+} | \mu_1, \dots, \mu_{r_-})}
\pm e^{\frac{i \pi}{2} (r_+ - r_-)} 
\varphi_{(\mu_1, \dots, \mu_{r_-} | \lambda_1, \dots, \lambda_{r_+})}
\right)
\end{equation}
\begin{equation}
C \varphi_{(\lambda_1, \dots, \lambda_{r_+} | \mu_1, \dots,
  \mu_{r_-})}^{\pm} = 
\pm \varphi_{(\lambda_1, \dots, \lambda_{r_+} | \mu_1, \dots,
  \mu_{r_-})} ^{\pm}
\end{equation}
\\
\\
\emph{Action of the shift operator.} The shift operator $e^{-iP}$ has 
a similar action on the Bethe states~:
\begin{equation}
e^{-iP} \varphi_{\alpha_1, \dots, \alpha_r} = 
\left[ \prod_{j=1}^{r} i e^{-ip_1(\alpha_j+i\pi)} \right] \ 
\varphi_{\alpha_1+i\pi, \dots, \alpha_r+i\pi}
\end{equation}
where~:
\begin{equation}
\exp \left[ i p_1(\alpha) \right] = 
\frac{\sinh \frac{1}{2} (\alpha + i \gamma) }
{\sinh \frac{1}{2} (\alpha - i \gamma) }
\end{equation}

\subsection{Eigenvalues}
The eigenvalue of the transfer matrix $\mathcal{T}(u)$ associated 
to the Bethe sta\-te $\varphi_{\alpha_1, \dots, \alpha_r}$ is~:
\begin{equation}
\Lambda(\alpha_1, \dots, \alpha_r | u) = 
\mu(\alpha_1, \dots, \alpha_r | u) 
         \mu(\alpha_1+i\pi, \dots, \alpha_r+i\pi | u)
\end{equation}
where
%\begin{eqnarray}
% & & \mu(\alpha_1, \dots, \alpha_r | u) = 
%\left( i/2 \right )^N \times \notag \\
% & & \left\{
%\left[ \sin 2(u-\gamma) \right]^N 
%\prod_{j=1}^{r} e^{-i p_1(\alpha_j - 2iu)} 
%+ 
%\left( \sin 2u \right)^N 
%\prod_{j=1}^{r} e^{i p_1(\alpha_j - 2iu + 2i \gamma)} \right\} 
%\end{eqnarray}
\begin{eqnarray}
 & & \mu(\alpha_1, \dots, \alpha_r | u) \equiv
\left( i/2 \right)^N \times \notag \\
 & & \left\{
\left[ \sin 2(u-\gamma) \right]^N 
\prod_{j=1}^{r} \frac{\sinh \frac{1}{2}(\alpha_j-2iu-i\gamma)}
                   {\sinh \frac{1}{2}(\alpha_j-2iu+i\gamma)}
+
\left( \sin 2u \right)^N 
\prod_{j=1}^{r} \frac{\sinh \frac{1}{2}(\alpha_j-2iu+3i\gamma)}
                   {\sinh \frac{1}{2}(\alpha_j-2iu+i\gamma)}
\right\} \notag \\
\end{eqnarray}
In the limit $u \to 0$, the energy of the state 
$\varphi_{\alpha_1, \dots, \alpha_r}$ is defined as~:
\begin{eqnarray}
\mathcal{E}(\alpha_1, \dots, \alpha_r) & = & \frac{1}{2} \sin 2 \gamma \ 
\frac{\partial \log \Lambda}{\partial u} (\alpha_1, \dots, \alpha_r | 0) \\
\ & = & -2N \cos 2 \gamma + \sum_{j=1}^{r} 
\frac{2 \sin^2 2 \gamma}{\cosh 2 \alpha_j - \cos 2 \gamma}
\end{eqnarray}
By construction, the state $\varphi_{\alpha_1, \dots, \alpha_r}$ is an
eigenvector of the Hamiltonian $\mathcal{H}$, with eigenvalue
$E(\alpha_1, \dots, \alpha_r)$.

\subsection{Reminder : the homogeneous six-vertex model}

Consider the homogeneous six-vertex model defined by the $R_0(u_0)$ 
matrix given in equation~\eqref{eq:R-matrix-6v}, with parameter 
$\gamma_0$, on a lattice of width $N$.

In the anisotropic limit $u_0 \to 0$, the logarithmic derivative
of the transfer matrix $T_0(u_0)$ is equal to the XXZ Hamiltonian~:
\begin{eqnarray} 
H_0 & \equiv & -\sin \gamma_0  \left[ T_0(0) \right]^{-1} 
\frac{\partial T_0}{\partial u_0} (0) \\
 & = & \sum_{m=1}^{N} ( \cos \gamma_0 - E^{0}_{m} ) \label{eq:HXXZ-1}
\end{eqnarray}
where the operator~$E^{0}_{m}$ is a generator of the Temperley-Lieb
algebra with parameter $\gamma_0$, 
like in equation~\eqref{eq:TL-Pauli-1}:
\begin{equation}
E^{0}_{m} = \frac{1}{2} \left[
\sigma_m^x \ \sigma_{m+1}^x + \sigma_m^y \ \sigma_{m+1}^y
- \cos \gamma_0 (\sigma_m^z \ \sigma_{m+1}^z - \myid) 
- i \sin \gamma_0 (\sigma_m^z - \sigma_{m+1}^z ) \right]
\end{equation}

On a lattice with periodic boundary conditions, the system is 
solvable by Bethe ansatz. The Bethe equations for $r$ particles 
are~:
\begin{equation} \label{eq:BE-6v}
\forall j \in \{1, \dots r\} \qquad
\exp \left[ i N k_0(\lambda_j) \right] = 
- \prod_{l=1}^r \exp \left[ -i \phi_0(\lambda_j, \lambda_l) \right]
\end{equation}
where the one-particle momentum and the scattering amplitude are~:
\begin{equation} \label{eq:k0}
\exp \left[ i k_0(\lambda) \right] = \frac{\sinh (\frac{i}{2} \gamma_0 - \lambda )}
                         {\sinh (\frac{i}{2} \gamma_0 + \lambda )} ,
\end{equation}
\begin{equation} \label{eq:phi0}
\exp \left[ i \phi_0(\lambda, \lambda') \right] = 
\frac{\sinh (\lambda - \lambda' + i \gamma_0 )}
     {\sinh (\lambda - \lambda' - i \gamma_0 )}
\end{equation}
The associated eigenvalue of the transfer matrix $T_0(u_0)$ is~:
\begin{eqnarray}
\Lambda_0(\lambda_1, \dots, \lambda_r | u_0) = & & 
\left[ \sin(\gamma_0-u_0)\right]^N 
  \prod_{j=1}^{r} \left( 
- \frac{\sinh(\lambda_j + i u_0 + \frac{i}{2} \gamma_0)}
       {\sinh(\lambda_j + i u_0 - \frac{i}{2} \gamma_0)}
     \right) \notag \\
& & + \left( \sin u_0 \right)^N 
  \prod_{j=1}^{r} \left( 
- \frac{\sinh(\lambda_j + i u_0 - \frac{3i}{2} \gamma_0)}
       {\sinh(\lambda_j + i u_0 - \frac{i}{2} \gamma_0)}
     \right) \notag \\
\label{eq:lambda0}
\end{eqnarray}

\subsection{Common eigenvalues between the staggered 
  and homogeneous six-vertex models}
\label{sec:common}

If the parameters of the staggered and the homogeneous 
six-vertex models are related by~:
\begin{eqnarray}
\gamma_0 &=& \pi - 2 \gamma \\
u_0 &=& -2 u
\end{eqnarray}
then one has the relations~:
\begin{equation} \label{eq:k-k0}
\exp \left[ 2i \tilde{k}(\lambda) \right] = \exp \left[ i k_0(\lambda) \right]
\end{equation}
\begin{equation} \label{eq:phi-phi0}
\phi_1(\lambda, \lambda') + \phi_{-1} (\lambda, \lambda') = \phi_0 (\lambda, \lambda')
\end{equation}
These relations suggest that the states~:
\begin{equation} \label{eq:XXZ-state}
\varphi_{(\lambda_1, \dots, \lambda_r|\lambda_1, \dots, \lambda_r)}
\end{equation}
have properties described by the homogeneous six-vertex model.
The states~\eqref{eq:XXZ-state} are obtained as the solution of the
Bethe equations~\eqref{eq:BE-H-4}--\eqref{eq:BE-H-5} when the Bethe
integers on the two lines are the same~:
\begin{equation} \label{eq:XXZ-state-1}
r_+ = r_- = r
\end{equation}
\begin{equation} \label{eq:XXZ-state-2}
\forall j \in \{ 1, \dots, r \} \ , \qquad I^+_j = I^-_j = I_j
\end{equation}
As a consequence of relations~\eqref{eq:k-k0}--\eqref{eq:phi-phi0}, the
Bethe equations~\eqref{eq:BE-H-4}--\eqref{eq:BE-H-5} are equivalent
to the Bethe equation~\eqref{eq:BE-6v} for the homogeneous six-vertex 
model. Moreover, the eigenvalue of the
transfer matrix $T(u)$ can be written in terms of the 
eigenvalue~\eqref{eq:lambda0}~:
\begin{equation}
\Lambda \left( \lambda_1+i\frac{\pi}{2}, \lambda_1-i\frac{\pi}{2}, \dots, 
\lambda_r+i\frac{\pi}{2}, \lambda_r-i\frac{\pi}{2} | u \right) = 
(-1/4)^N \left[ \Lambda_0(\lambda_1, \dots, \lambda_r | u_0) \right]^2
\end{equation}
As a consequence, in the anisotropic limit, the energies
$\mathcal{E}$, $\mathcal{E}_0$ of the Hamiltonians $\mathcal{H}$,
$H_0$ are related by~:
\begin{equation} \label{eq:E-E0}
\mathcal{E} \left( \lambda_1+i\frac{\pi}{2}, \lambda_1-i\frac{\pi}{2}, \dots, 
\lambda_r+i\frac{\pi}{2}, \lambda_r-i\frac{\pi}{2} \right) = 
2 \mathcal{E}_0(\lambda_1, \dots, \lambda_r)
\end{equation}
The total momenta are equal~:
\begin{equation} \label{eq:ktot-k0tot}
k_{\mathrm{tot}} = k_{0, \mathrm{tot}}
\end{equation}

\section{Finite-size study of the Bethe equations}

\subsection{Reminder : the XXZ spin-chain Hamiltonian.}

\subsubsection{Definition}

The XXZ Hamiltonian~\eqref{eq:HXXZ-1} on a lattice of width $N$
can be written~:
\begin{equation} \label{eq:HXXZ-2}
H_0 = -1/2 \ \sum_{m=1}^{N} \left[ 
\sigma_m^x \sigma_{m+1}^x + \sigma_m^y \sigma_{m+1}^y 
-\cos \gamma_0 \ ( \sigma_m^z \sigma_{m+1}^z + \myid )
\right]
\end{equation}
\begin{equation}
\qquad 0 \leq \gamma_0 \leq \pi
\end{equation}
with periodic boundary conditions~:
\begin{equation} \label{eq:PBC}
\sigma_{N+1}^{\mu} = \sigma_1^{\mu}
\end{equation}

\subsubsection{Ground state}
The ground state of the Hamiltonian \eqref{eq:HXXZ-2} is 
given by the symmetric ``half-filled Fermi sea'' in the sector with
$r=N/2$ particles.
Consider one of the particle-hole excitations around the Fermi momenta $\pm k_F$~:
\begin{equation} \label{eq:part-hole-1}
k_F-k_0/2 \to k_F+k_0/2
\end{equation}
\begin{equation} \label{eq:part-hole-2}
-k_F+k_0/2 \to -k_F-k_0/2
\end{equation}
These excited states have total momentum $\pm k_0$. In the
thermodynamic limit $N \to \infty$ with the ratio $r/N$ fixed,
the Bethe equations~\eqref{eq:BE-6v} give a linear integral 
equation for the density of particles. Solving this equation
gives access to the dispersion relation of the excitations 
\eqref{eq:part-hole-1}--\eqref{eq:part-hole-2}~:
\begin{equation} \label{eq:v0}
\mathcal{E}_0(k_0) 
\simeq v_0 |k_0|, \qquad v_0=\frac{\pi \sin \gamma_0} {\gamma_0}
\end{equation}
This shows that, in the thermodynamic limit, the spectrum has no 
finite gap above the ground state. This is an indication 
that the equivalent (1+1)-dimen\-sional quantum field theory is 
conformally invariant. The velocity $v_0$ appears in the finite-size 
determination of the central charge~:
\begin{equation}
\mathcal{E}_{0,\mathrm{gr}}(N) = N \ e_{0,\infty} - \frac{\pi v_0 c}{6 N} + o(N^{-1})
\end{equation}
The central charge of the Hamiltonian $H_0$ is $c=1$. The energy
density per site is~:
\begin{equation} \label{eq:e-infinity}
e_{0,\infty} = \cos \gamma_0 - 2 \sin^2 \gamma_0 \int_0^\infty
\frac{dx}{\cosh(\pi x) [\cosh (2 \gamma_0 x) - \cos \gamma_0]}
\end{equation}

\subsubsection{Low-lying excitations}
A class of Bethe states are low-lying excitations, with energies~:
\begin{equation}
\mathcal{E}_0(N) = \mathcal{E}_{0,\mathrm{gr}}(N) + \frac{2 \pi v_0 x}{N} + o(N^{-1})
\end{equation}
where $x$ is called the physical exponent.
We describe the primary states, and then their descendants.
In the sector with $r=N/2-n$ particles, the lowest-energy state is
given by the distribution of the Bethe integers which
is symmetric around zero.
In the same sector, denote by $\{ n, m \}$ the state
obtained after $m$ backscatterings from the left Fermi level
to the right Fermi level~:
\begin{equation} \label{eq:backscattering}
I_1, \dots I_r =-\frac{r-1}{2}+m, \dots \frac{r-1}{2}+m
\end{equation}
The physical exponents of this class of states are~:
\begin{equation}
x_{n, m}^{(0)} = n^2 \frac{g}{2} + m^2 \frac{1}{2g}, 
\qquad g= \frac{\pi - \gamma_0}{\pi}
\end{equation}
Descendant states are obtained by performing particle-hole excitations
on the above states. For example, starting from the ground state
distribution and setting $I_r$ to $(r+1)/2$, one obtains
a state with physical exponent $x=1$.

\subsection{Ground state of the Hamiltonian $\mathcal{H}$}

In this section, the ground state of the Hamiltonian~\eqref{eq:h},
corresponding to the anisotropic limit of the staggered six-vertex model,
is discussed. The system width $N$ is assumed to be even. The case
of an odd system width $N$ will be discussed in 
section~\ref{sec:N-odd}.
The ground state of the Hamiltonian $\mathcal{H}$ is
given by the symmetric distribution of the Bethe integers
in the sector $r_+ = r_- = N/2$~:
\begin{eqnarray}
I^+_1, \dots I^+_{N/2} &=& -\frac{N/2-1}{2}, -\frac{N/2-1}{2}+1, \dots, \frac{N/2-1}{2} \\
I^-_1, \dots I^-_{N/2} &=& -\frac{N/2-1}{2}, -\frac{N/2-1}{2}+1, \dots, \frac{N/2-1}{2}
\end{eqnarray}
Note that this state has twice the energy of the ground state of 
$H_{\mathrm{XXZ}}$.

Consider the double particle-hole excitation obtained by
performing the transformation~\eqref{eq:part-hole-1} on both
Fermi seas. This excitation is of the 
type~\eqref{eq:XXZ-state-1}--\eqref{eq:XXZ-state-2}, and therefore,
using equations~\eqref{eq:v0},
\eqref{eq:ktot-k0tot} and~\eqref{eq:E-E0} its momentum and energy are~:
\begin{equation}
k = k_0
\end{equation}
\begin{equation} \label{eq:v-v0}
\mathcal{E}(k) \simeq 2 \ v_0 \ | k_0 |
\end{equation}
Thus, the rapidity of the particle-hole excitations around the
ground state of the Hamiltonian $\mathcal{H}$ is~:
\begin{equation}
v = 2 \ v_0
\end{equation}
Compare the asymptotic behaviors of the finite-size ground state
energies for Hamiltonians $H_0$, $\mathcal{H}$~:
\begin{equation}
\mathcal{E}_{0, \mathrm{gr}} (N) = N e_{0,\infty} - \frac{\pi v_0}{6 N} + o(N^{-2})
\end{equation}
\begin{equation}
\mathcal{E}_{\mathrm{gr}} (2N) = 2N e_\infty - \frac{\pi c v}{6\times 2N} + o(N^{-2})
\end{equation}
As a consequence of the identities \eqref{eq:E-E0} and
\eqref{eq:v-v0}, the central charge of Hamiltonian $\mathcal{H}$ is
$c=2$. The energy density per site of Hamiltonian $\mathcal{H}$ is~:
\begin{equation}
e_\infty = e_{0,\infty}
\end{equation}

\subsection{Low-energy spectrum of the Hamiltonian $\mathcal{H}$}
\label{sec:spectrum-H}

\subsubsection{Bethe excitations}

The excitations considered are combinations of particle and
backscattering excitations on the Bethe integers $I^{+}_j, I^{-}_j$.
Denote $\varphi_{(n_+, n_-), (m_+, m_-)}$
the Bethe state defined by the numbers of particles~:
\begin{eqnarray}
r_+ &=& N/2 - n_+ \label{eq:n-plus} \\
r_- &=& N/2 - n_- \label{eq:n-minus} 
\end{eqnarray}
and the Bethe integers~:
\begin{eqnarray}
I^+_1, \dots I^+_{r_+} &=& -\frac{r_+ - 1}{2}+m_+, \dots ,
\frac{r_+ - 1}{2}+m_+ \label{eq:I-plus} \\
I^-_1, \dots I^-_{r_-} &=& -\frac{r_- - 1}{2}+m_-, \dots ,
\frac{r_- - 1}{2}+m_- \label{eq:I-minus} 
\end{eqnarray}
The states described in section~\ref{sec:common} are of the type 
$\varphi_{(n,n), (m,m)}$. Their physical exponents are~:
\begin{equation}
x_{ (n, n), (m, m) } =
2 \ x_{n, m}^{(0)} = n^2 g + m^2 \frac{1}{g}
\end{equation}
where
\begin{equation} \label{eq:g}
g=2 \gamma / \pi
\end{equation}
The analytical study \cite{jacobsen-saleur06} of the thermodynamic
limit of the Bethe equations suggests the following general form for
the physical exponents~:
\begin{eqnarray} \label{eq:bethe-spectrum}
x_{ (n_+, n_-), (m_+, m_-) } = & \ &
\frac{1}{4} g (n_+ + n_-)^2
+ \frac{1}{4g} (m_+ + m_-)^2 \notag \\
\ & + & \frac{1}{4} K(\gamma, N) (n_+ - n_-)^2
+ \frac{1}{4K(\gamma, N)} (m_+ - m_-)^2 \notag \\
\ & \ & 
\end{eqnarray}
where the coupling constant $K(\gamma, N)$ tends to zero
as $N$ goes to infinity. One of the key questions is to determine the 
way this constant actually vanishes.

\subsubsection{General structure of the low-energy spectrum}

Note that the states with $m_+ \neq m_-$ 
are not part of the low-energy spectrum. Thus, in the following, the
discussion will concern only the states with 
$m_+ = m_- = m$.
For these states, the total magnetization and the total momentum
read~:
\begin{eqnarray}
S &=& n_+ + n_- \\
k_{\mathrm{tot}} &=& \frac{\pi}{N} m (N - n_+ - n_-)
\end{eqnarray}

\begin{figure}
\begin{center}
\includegraphics[scale=0.6]{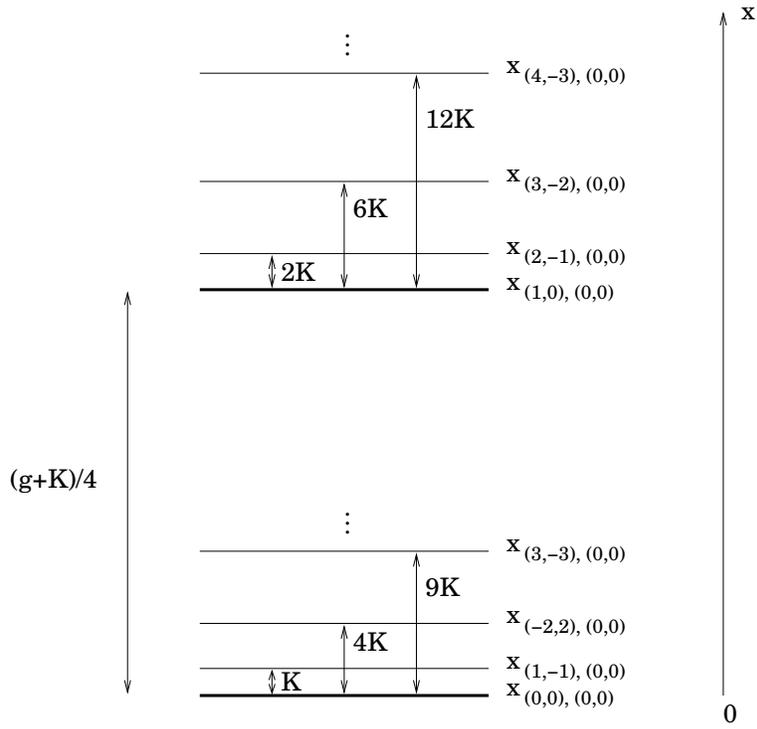}
\caption{The first levels of the fundamental sector $(n_+ + n_-= 0, m=0)$ and the
sector defined by $(n_+ + n_-= 1, m=0)$. The floor states
(including the ground state) are represented by bold lines.
The energies are rescaled as physical exponents.}
\label{fig:spectrum}
\end{center} 
\end{figure}

Equation~\eqref{eq:bethe-spectrum} determines the structure of the
low-energy spectrum. The quantities $(n_+ + n_-)$ and $m$ define a
sector of given total spin $S$ and total momentum $k_{\mathrm{tot}}$.
We call ``floor states'' the lowest-energy states of these sectors.
These are the states $\varphi_{(n, n),(m,m)}$ and
$\varphi^{\pm}_{(n+1, n),(m,m)}$, where $n,m$ are integers.
The physical exponents of the floor states are finite in the limit
$N \to \infty$~:
\begin{eqnarray} 
x_{(n, n),(m,m)} &=& g n^2 + \frac{m^2}{g} \label{eq:x-floor-even} \\
x_{(n+1, n),(m,m)} &=& \frac{g}{4} (2n+1)^2 + \frac{m^2}{g} +
\frac{K(\gamma,N)}{4} \label{eq:x-floor-odd}
\end{eqnarray}
Note that the states $\varphi_{(n, n),(m,m)}$
are those described in section~\ref{sec:common}.
Within each sector, higher energy states are obtained, 
starting from the floor
state, by ``moving'' $n'$ particles from the line 
$\mathrm{Im} (\alpha) = \pi/2$ to the line 
$\mathrm{Im} (\alpha) = -\pi/2$ (or the reverse).
The physical exponent of the resulting state differs from the
floor exponent by a quantity proportional to the coupling
constant $K(\gamma,N)$~:
\begin{eqnarray} 
x_{(n+n', n-n'),(m,m)} &=& x_{(n, n),(m,m)} + (n')^2 K(\gamma, N)
\label{eq:xm-even} \\
x_{(n+n'+1, n-n'),(m,m)} &=& x_{(n+1, n),(m,m)} + n'(n'+1) K(\gamma,N) 
\label{eq:xm-odd}
\end{eqnarray}
Thus, if the constant $K$ indeed vanishes in the limit $N \to \infty$,
the gaps between the floor state and the higher states in the
sector should also vanish. The structure of the
spectrum is illustrated in figure~\ref{fig:spectrum}.

Numerical calculations are used to confirm the 
form~\eqref{eq:bethe-spectrum} of the physical exponents,
and to determine the scaling law for $K(\gamma, N)$.

\subsubsection{Numerical calculation of the physical exponents}

\begin{figure}
\begin{center}
\input{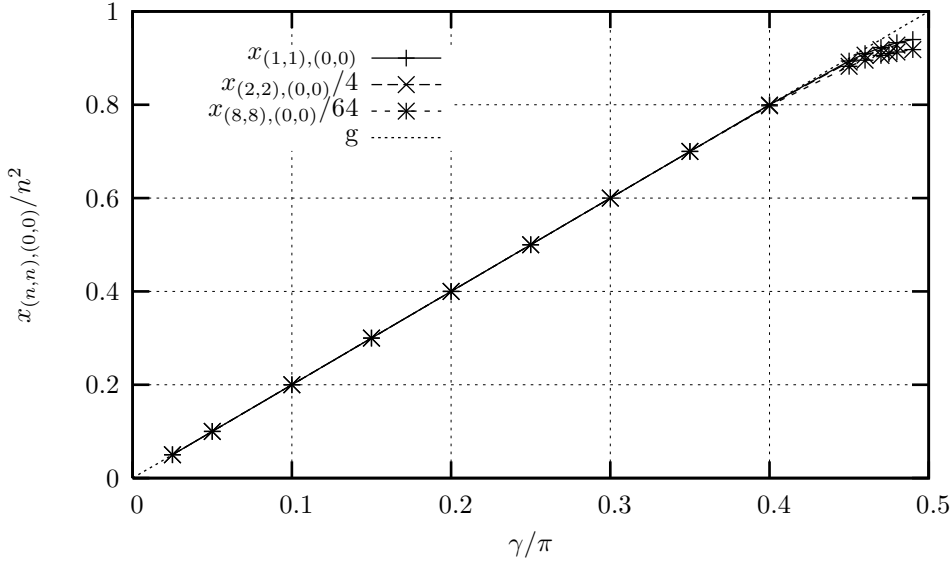}
\caption{The floor exponents $x_{(n,n),(0,0)}$ as functions of
  $\gamma$, for a system width $N=1024$.
  The expected values are $x_{(n,n),(0,0)} = g n^2$ 
  (equation~\eqref{eq:x-floor-even}).}
\label{fig:x.min.r.r}
\end{center} 
\end{figure}

\begin{figure}
\begin{center}
\input{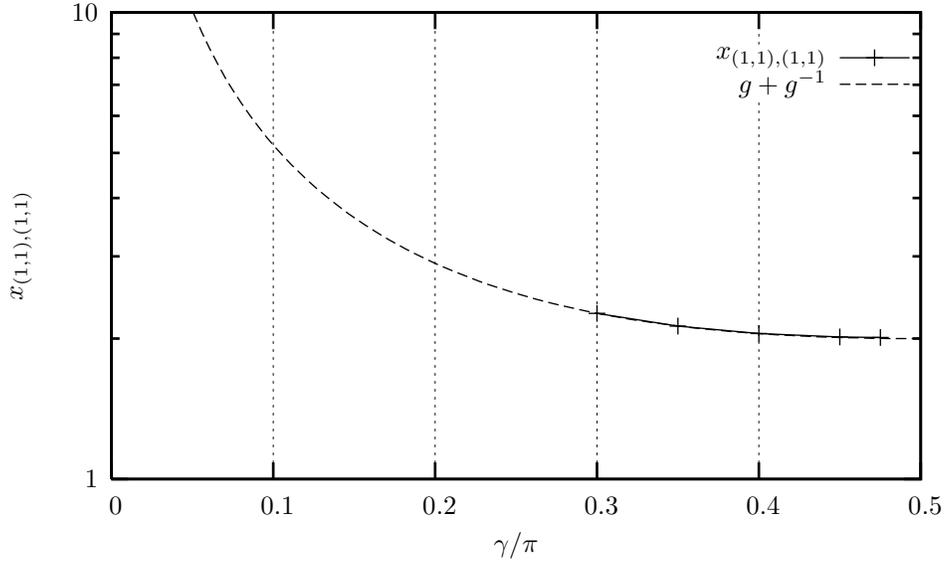}
\caption{The floor exponent $x_{(1,1),(1,1)}$ as a function of
$\gamma$, for a system width $N=512$. The expected value is
$x_{(1,1),(1,1)} = g + g^{-1}$ (equation~\eqref{eq:x-floor-even}).}
\label{fig:x.bs1.1.1.1}
\end{center} 
\end{figure}

\begin{figure}
\begin{center}
\input{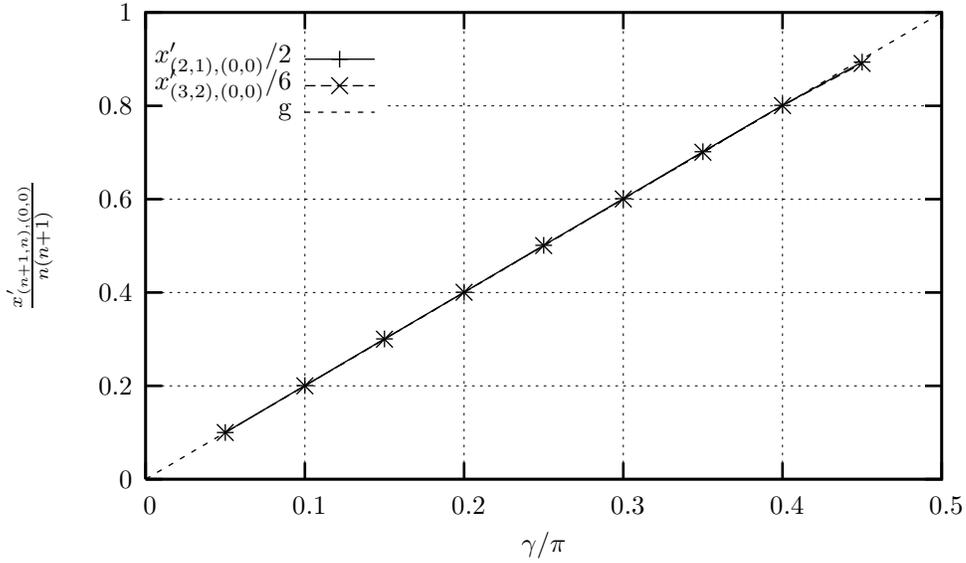}
\caption{The floor exponent $x'_{(n+1,n),(0,0)} \equiv
  x_{(n+1,n),(0,0)} - x_{(1,0),(0,0)}$ as a function of
$\gamma$, for a system width $N=1024$. The expected value is
$x'_{(n+1,n),(0,0)}= g n(n+1)$ (equation~\eqref{eq:x-floor-odd}).}
\label{fig:xp.min.r+1.r}
\end{center} 
\end{figure}

\begin{figure}
\begin{center}
\input{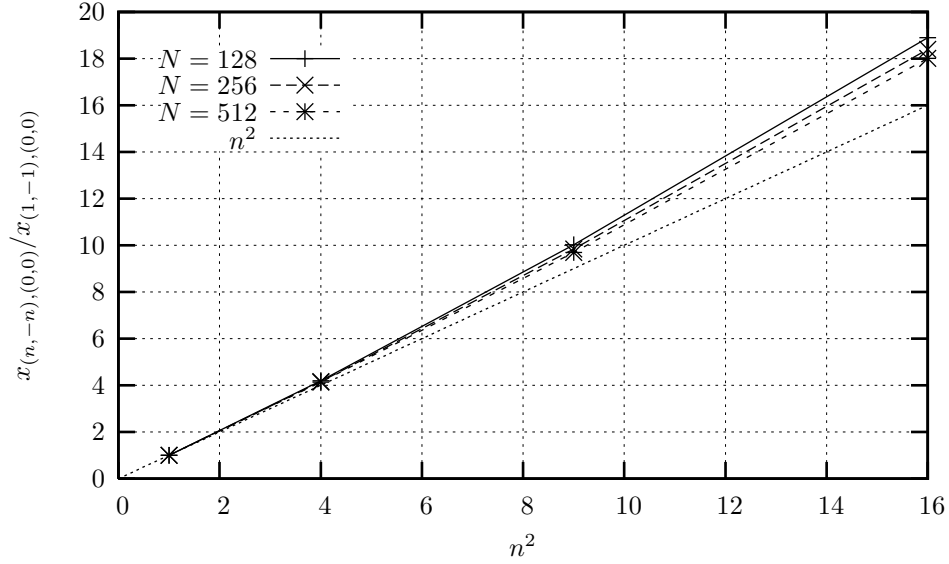}
\caption{The exponent $x_{(n,-n),(0,0)}$ as a function of $n$,
for parameter $\gamma=0.3 \pi$. The expected values are
$x_{(n,-n),(0,0)} = K(\gamma,N) \ n^2$ (equation~\eqref{eq:xm-even}).}
\label{fig:x.min.r.-r}
\end{center} 
\end{figure}

\begin{figure}
\begin{center}
\input{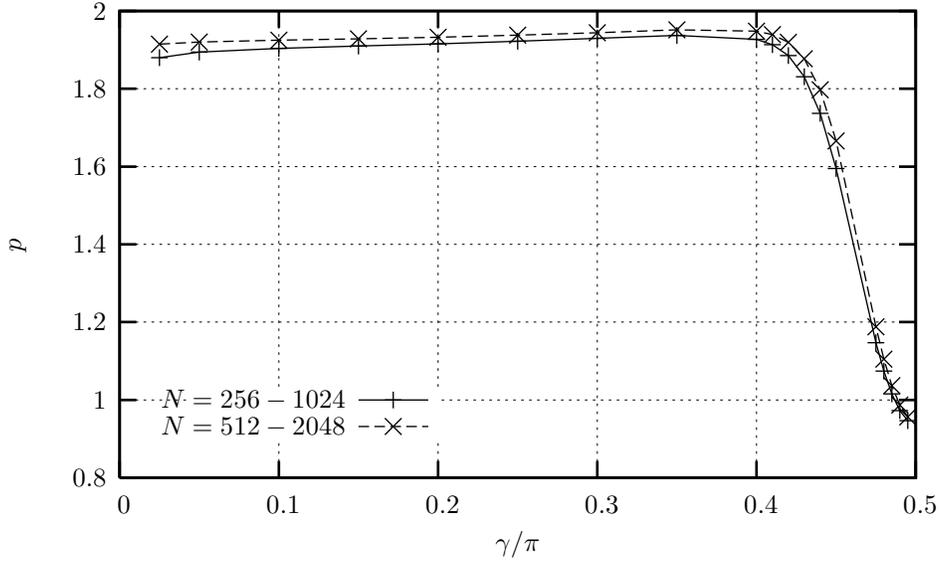}
\caption{Estimation of the exponent $p$ in the scaling
  law~\eqref{eq:K-scaling-p}. The estimator for $p$ is obtained
by eliminating the unknowns $A,B$ from the equations
$\log K(\gamma,N) = \log A - p \log (B + \log N)$, for three system widths.
The estimators converge slowly to $p=2$ for $\gamma<\pi/2$, and
to $p=1$ for $\gamma=\pi/2$.
}
\label{fig:p.min1.-1}
\end{center} 
\end{figure}

\textit{Numerical procedure.} 
When the Bethe integers $I_j^\pm$ are fixed according to 
equations~\eqref{eq:I-plus}--\eqref{eq:I-minus},
the Bethe equations~\eqref{eq:BE-H-4}--\eqref{eq:BE-H-5} are a set of
non-linear equations for the variables $\lambda_j, \mu_j$.
These equations can be solved numerically, using the multidimensional
Newton-Raphson method \cite{nr92}.
The following starting point for the algorithm was found empirically
to lead to a good convergence~:
\begin{equation}
\lambda_j^{\mathrm{init}} = \mathrm{Arctanh} \left[
\frac{\tan(\pi I_j^+ / N)}{U}
\right]
\end{equation}
where the additional parameter is set to $U=10$.
The exponents are then estimated, relatively to the ground state,
or to the floor state, using~:
\begin{eqnarray} 
\mathcal{E}_a - \mathcal{E}_{\mathrm{gr}} &\sim &
\frac{2 \pi v}{2N} x_a \\
\mathcal{E}_a - \mathcal{E}_b &\sim &
\frac{2 \pi v}{2N} (x_a-x_b)
\end{eqnarray}
where $a,b$ denote any states of the spectrum.
\\
\\
\textit{Results for the floor exponents.}
These are shown in
figures~\ref{fig:x.min.r.r}--\ref{fig:xp.min.r+1.r}.
These results agree with the form~\eqref{eq:bethe-spectrum}.
\\
\\
\textit{Results for the gaps inside a given sector.}
These gaps are given in
equations~\eqref{eq:xm-even}-\eqref{eq:xm-odd},
as a function of the integer $n'$ and the coupling constant
$K(\gamma, N)$. The first step is to check the dependence on
$n'$. See figures~\ref{fig:x.min.r.-r}--\ref{fig:xm.min.1+r.-r}.
\\
\\
\textit{Results for the coupling constant.}
Equations~\eqref{eq:xm-even}--\eqref{eq:xm-odd} relate
the gaps within a given sector to the coupling constant
$K(\gamma, N)$. These are used to determine the scaling law
of this quantity. Theoretical arguments (see
section~\ref{sec:sigma}) predict the following form~:
\begin{equation} \label{eq:K-scaling-p}
K \simeq
\frac{A}{\left[ B + \log N \right]^p}
\end{equation}
where the exponent $p$ may take the values $p=1,2$.
The exponent $p$ is estimated numerically, assuming
that $K$ behaves as equation~\eqref{eq:K-scaling-p} 
(see figure~\ref{fig:p.min1.-1}).
A crossover is observed between the behaviours~:
\begin{eqnarray}
K(\gamma, N) &\simeq&
\frac{A(\gamma)}{\left[ B(\gamma) + \log N \right]^2}
\qquad (\gamma < \pi/2) \label{eq:K-scaling} \\
K(\pi/2, N)  &\simeq&
\frac{A'}{B' + \log N} \label{eq:K-scaling-pi-2}
\end{eqnarray}

\begin{figure}
\begin{center}
\input{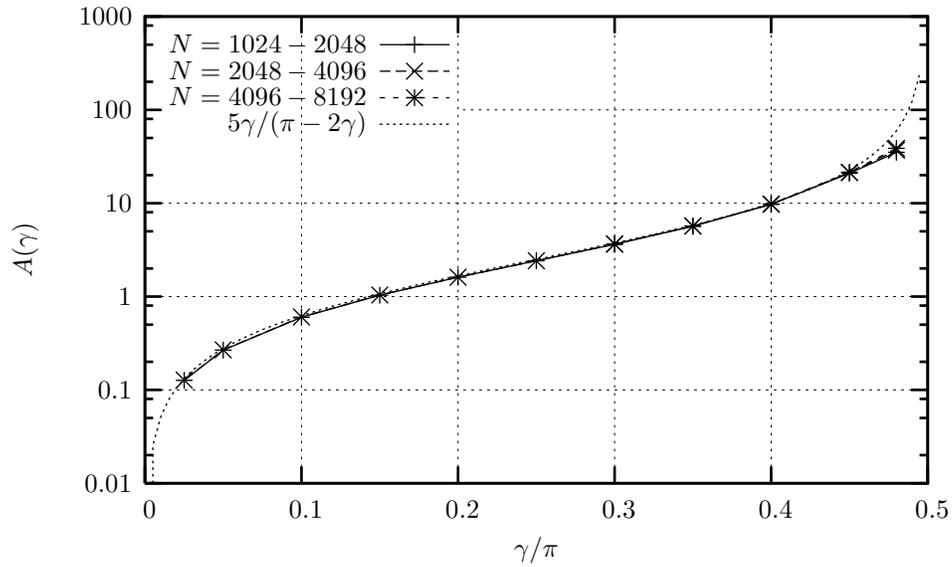}
\caption{Determination of the factor $A(\gamma)$ in the scaling 
law~\eqref{eq:K-scaling}, using a numerical estimation
of exponent $x_{(1,-1),(0,0)}$.}
\label{fig:a.min1.-1}
\end{center} 
\end{figure}

The factor $A(\gamma)$ in the scaling law~\eqref{eq:K-scaling} is 
estimated numerically by eliminating the
term $B(\gamma)$ between two system widths. 
The finite-size estimators are defined as~:
\begin{equation}
A(\gamma, N_1, N_2) \equiv \left\{
\frac{ \log(N_1/N_2) }{[K(\gamma, N_1)]^{-1/2} - [K(\gamma, N_2)]^{-1/2}}
\right\}^2
\end{equation}
The numerical results (see figure~\ref{fig:a.min1.-1}) 
allow us to conjecture the following form for the factor $A(\gamma)$~:
\begin{equation}
A(\gamma) = \frac{5 \gamma}{\pi - 2 \gamma}
\end{equation}
The correction term $B(\gamma)$ depends on the sector considered.

\begin{figure}
\begin{center}
\input{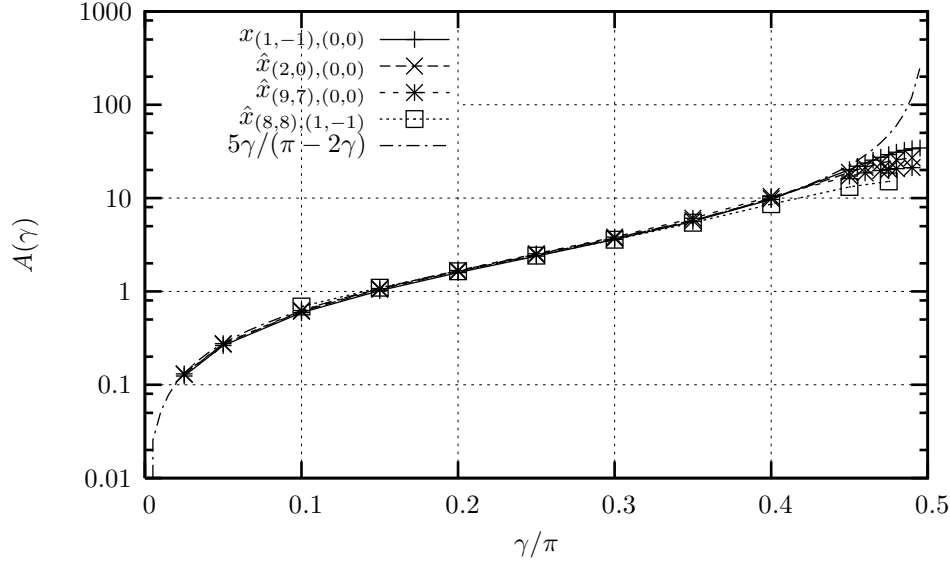}
\caption{Various determinations of the factor $A(\gamma)$, for system
  widths $N=256-512$.}
\label{fig:a.various}
\end{center} 
\end{figure}

Each sector determines a specific scaling law for $K(\gamma, N)$,
and these determinations can be compared.
In addition, the coupling constant $K(\gamma, N)$ appears in the
expression of ``infinite physical exponents''~:
\begin{equation}
x_{(n,n),(1,-1)} - x_{(n,n),(0,0)} = \frac{1}{K(\gamma, N)}
\end{equation}
The results are shown in figure~\ref{fig:a.various}.

\begin{figure}
\begin{center}
\input{xm.min.1+r.1-r}
\caption{The exponent 
$\hat{x}_{(1+n,1-n),(0,0)} \equiv x_{(1+n,1-n),(0,0)} - x_{(1,1),(0,0)}$ 
as a function of $n$, for parameter $\gamma=0.3 \pi$. 
The expected values are
$\hat{x}_{(1+n,1-n),(0,0)} = K(\gamma,N) \ n^2$ (equation~\eqref{eq:xm-even}).}
\label{fig:xm.min.1+r.1-r}
\end{center} 
\end{figure}

\begin{figure}
\begin{center}
\input{xm.bs.1+r.1-r.1.1}
\caption{The exponent 
$\hat{x}_{(1+n,1-n),(1,1)} \equiv x_{(1+n,1-n),(1,1)} -
x_{(1,1),(1,1)}$ 
as a function of $n$, for parameter $\gamma=0.4 \pi$.
The expected values are
$\hat{x}_{(1+n,1-n),(1,1)} = K(\gamma,N) \ n^2$ (equation~\eqref{eq:xm-even}).}
\label{fig:xm.bs.1+r.1-r.1.1}
\end{center} 
\end{figure}
\begin{figure}

\begin{center}
\input{xm.min.1+r.-r}
\caption{The exponent 
$\hat{x}_{(1+n,-n),(0,0)} \equiv x_{(1+n,-n),(0,0)} -
x_{(1,0),(0,0)}$
 as a function of $n$, for parameter $\gamma=0.3 \pi$.
The expected values are
$\hat{x}_{(1+n,-n),(0,0)} = K(\gamma,N) \ n(n+1)$ (equation~\eqref{eq:xm-odd}).}
\label{fig:xm.min.1+r.-r}
\end{center} 
\end{figure}

\subsection{The case $N$ odd}
\label{sec:N-odd}

When the system width $N$ is odd, the structure of the spectrum
exhibits some differences from the case $N$ even.

The general formula for the finite-size corrections is identical to the formula for $N$ even~:
\begin{equation} \label{eq:E-general}
\mathcal{E}_{(n_+, n_-), (m_+, m_-)}(2N) =
2N e_{\infty} - \frac{\pi v}{6 \times 2N} \times 2 
+ \frac{ 2 \pi v}{2N} x_{(n_+, n_-), (m_+, m_-)} + o(N^{-1})
\end{equation}
where the exponent $x_{(n_+, n_-), (m_+, m_-)}$ 
is given by equation~\eqref{eq:bethe-spectrum}.
The subtlety is that equations~\eqref{eq:n-plus}--\eqref{eq:n-minus} imply that 
$n_+,n_-$ are half-odd integers.

\begin{figure}
\begin{center}
\input{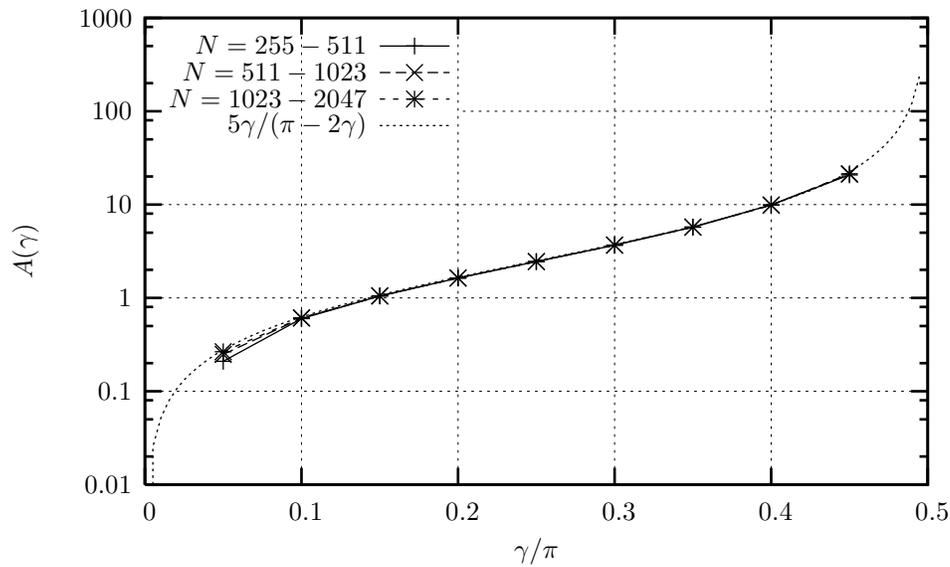}
\caption{Determination of the factor $A(\gamma)$ in the scaling 
law~\eqref{eq:K-scaling}
by the effective central charge $\tilde{c}$ for $N$ odd.}
\label{fig:a.min1.0.Nodd}
\end{center} 
\end{figure}

The lowest-energy states of the spectrum~\eqref{eq:E-general} are 
$\varphi^{\pm}_{(1/2, -1/2), (0,0)}$.
Note that these states belong to the sector $S=0$, 
and are the most closely packed to the imaginary $\alpha$ axis 
in this sector.
The energy of the ground state is~:
\begin{equation} \label{eq:E-gr-N-odd}
\mathcal{E}_{\mathrm{gr}}(2N) =
2N e_{\infty} - \frac{\pi v}{6 \times 2N} \times \tilde{c} + o(N^{-1})
\end{equation}
where the effective central charge is~:
\begin{equation}
\tilde{c} = 2 - 12 x_{(1/2, -1/2), (0,0)} \ , \qquad x_{(1/2, -1/2), (0,0)} = \frac{K(\gamma,N)}{4}
\end{equation}
The effective central charge is estimated numerically using 
equation~\eqref{eq:E-gr-N-odd} and 
the exact expression~\eqref{eq:e-infinity} 
of the energy density $e_{\infty}$ (see figure~\ref{fig:a.min1.0.Nodd}).
The scaling law for the coupling constant 
$K(\gamma,N)$ is consistent with equation~\eqref{eq:K-scaling}.
It is convenient to define the physical exponents of the excited
states with respect to the ground state~:
\begin{equation}
\tilde{x}_{(n_+, n_-), (m_+, m_-)} \equiv
x_{(n_+, n_-), (m_+, m_-)} - x_{(1/2, -1/2), (0,0)}
\end{equation}

The physical exponents of the floor states are~:
\begin{eqnarray}
\tilde{x}_{(n+1/2, n-1/2),(m,m)} &=& g n^2 + \frac{m^2}{g} 
\label{eq:x-floor-even-Nodd} \\
\tilde{x}_{(n-1/2, n-1/2),(m,m)} &=& \frac{g}{4} (2n-1)^2 
+\frac{m^2}{g} - \frac{K}{4} 
\label{eq:x-floor-odd-Nodd}
\end{eqnarray}
In particular, the states
$\varphi_{(-1/2, -1/2),(0,0)}, \varphi_{(1/2, 1/2),(0,0)}$ have
an energy which is twice that of the degenerate ground state 
of the XXZ spin-chain~\eqref{eq:HXXZ-2}
with an odd lattice width \cite{alcaraz88}.
These two states appear as excited states of the Hamiltonian $\mathcal{H}$.

The gaps inside a given sector are~:
\begin{eqnarray}
\tilde{x}_{(n+n'+1/2, n-n'-1/2),(m,m)} &=& \tilde{x}_{(n+1/2, n-1/2),(m,m)} + K n'(n'+1) 
\label{eq:xm-even-Nodd} \\
\tilde{x}_{(n+n'-1/2, n-n'-1/2),(m,m)} &=& \tilde{x}_{(n-1/2, n-1/2),(m,m)} +  K (n')^2
\label{eq:xm-odd-Nodd} 
\end{eqnarray}

\begin{figure}
\begin{center}
\input{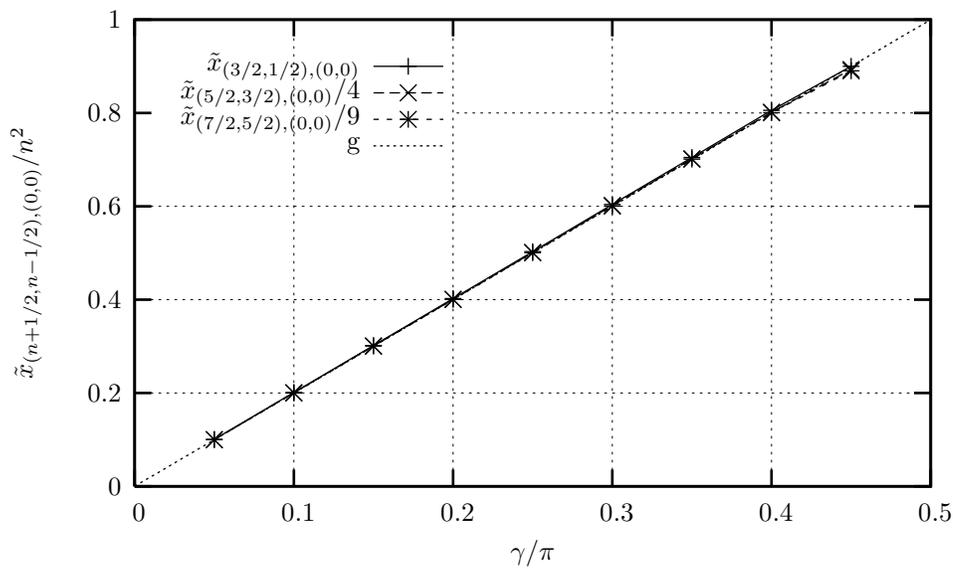}
\caption{The floor exponents $\tilde{x}_{(n+1/2, n-1/2),(0,0)}$ as functions
  of $\gamma$, for a lattice width $N=1023$. The expected values are
$\tilde{x}_{(n+1/2, n-1/2),(0,0)}= gn^2$ (equation \eqref{eq:x-floor-even-Nodd}).}
\label{fig:xt.min.r+1.r.Nodd}
\end{center} 
\end{figure}

\begin{figure}
\begin{center}
\input{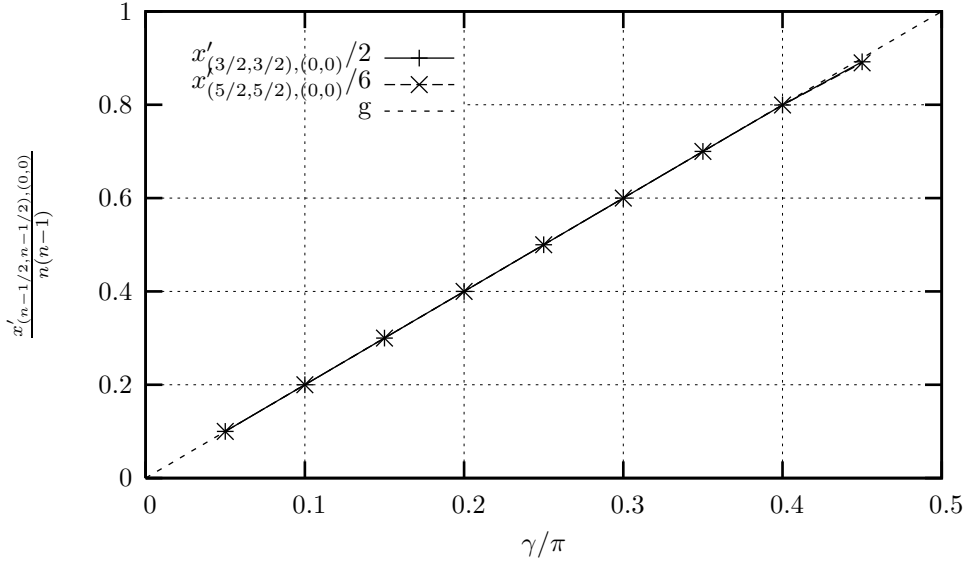}
\caption{The floor exponents 
$x'_{(n-1/2, n-1/2),(0,0)} \equiv \tilde{x}_{(n-1/2, n-1/2),(0,0)} - \tilde{x}_{(1/2, 1/2),(0,0)}$ 
as functions of $\gamma$, for a lattice width $N=1023$. The expected values are
$x'_{(n-1/2, n-1/2),(0,0)}= g n (n-1)$ (equation \eqref{eq:x-floor-odd-Nodd}).}
\label{fig:xp.min.r.r.Nodd}
\end{center} 
\end{figure}

\begin{figure}
\begin{center}
\input{xm.min.1+r.-r.Nodd}
\caption{The exponent 
$\hat{x}_{(1/2+n, -1/2-n),(0,0)} \equiv \tilde{x}_{(1/2+n, -1/2-n),(0,0)}-\tilde{x}_{(1/2, -1/2),(0,0)}$ 
as a function of $n$, for parameter $\gamma=0.3 \pi$. The expected values are
$\hat{x}_{(1/2+n, -1/2-n),(0,0)}= K(\gamma,N) \ n(n+1)$ (equation \eqref{eq:xm-even-Nodd}).}
\label{fig:xm.min.1+r.-r.Nodd}
\end{center} 
\end{figure}

\begin{figure}
\begin{center}
\input{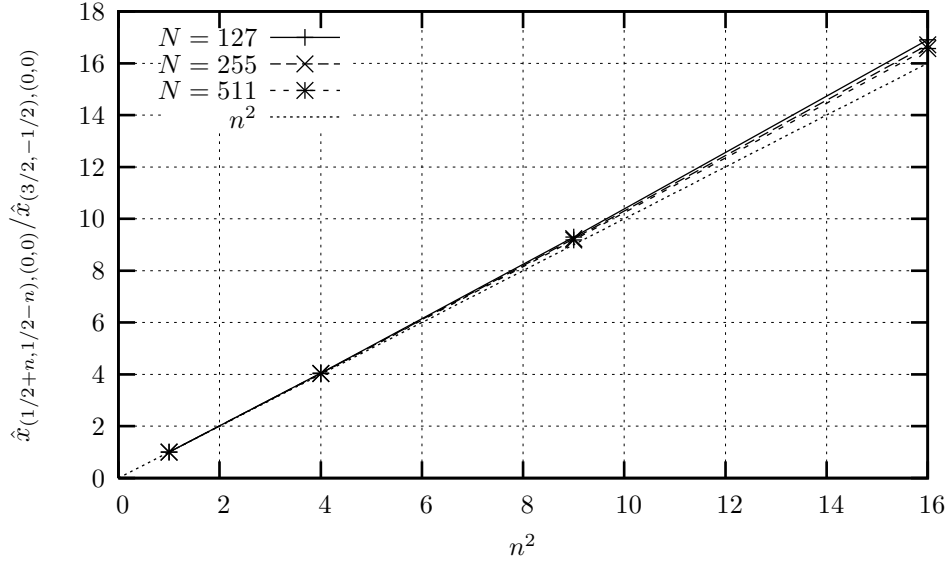}
\caption{The exponent 
$\hat{x}_{(1/2+n, 1/2-n),(0,0)} \equiv 
\tilde{x}_{(1/2+n, 1/2-n),(0,0)} - \tilde{x}_{(1/2, 1/2),(0,0)}$ 
as a function of $n$, for parameter $\gamma=0.3 \pi$.The expected values are
$\hat{x}_{(1/2+n, -1/2-n),(0,0)}= K(\gamma,N) \ n^2$ (equation \eqref{eq:xm-odd-Nodd}).}
\label{fig:xm.min.1+r.1-r.Nodd}
\end{center} 
\end{figure}

\section{Interpretation and relation to non-linear sigma models}
\label{sec:sigma}

As discussed in section 2, the limit $\gamma \to \pi / 2$ of the 
staggered six-vertex model coincides with a particular point of the 
$OSP(2|2)$ loop model of \cite{JRS}. This is a good starting point to 
understand the emergence of a continuous spectrum of critical exponents. 

\subsection{A reminder on intersecting loop models and Goldstone phases}

It turns out that the Mermin Wagner theorem forbidding the 
spontaneous breaking of a continuous symmetry in two dimensions does 
not hold for supergroups (because of the lack of unitarity), and that 
models with orthosymplectic $OSP(m|2n)$ symmetry do  exhibit a low 
temperature phase with spontaneous broken symmetry provided $m-2n<2$. 
More precisely, consider the non linear sigma model with target space 
the supersphere $S^{m,2n}=OSP(m|2n)/OSP(m-1|2n)$, a ``supersymmetric''
extension of the usual $O(N)$ sigma model. Use as coordinates a real 
scalar field~:
\begin{equation}
\phi\equiv (\phi_{1},\ldots,\phi_{m},\psi_{1},\ldots,\psi_{2n})
\end{equation}
and the invariant bilinear form
\begin{equation}
\phi \cdot \phi'=\sum \phi_{a}\phi'_{a}+\sum 
J_{\alpha\beta}\psi_{\alpha}\psi'_{\beta}
\end{equation}
where $J_{\alpha\beta}$ is the symplectic form which we take 
consisting of diagonal blocks~:
$\left(\begin{array}{cc} 
    0&1\\
    -1&0\end{array}\right)$. 
The unit supersphere is defined by the constraint~:
\begin{equation}
  \phi \cdot \phi=1
\end{equation}
The action of the sigma model (conventions are that the Boltzmann 
weight is $e^{-S}$) reads~:
\begin{equation}
  S=\frac{1}{2g_{\sigma}}\int d^{2}x \partial_{\mu}\phi \cdot \partial_{\mu}\phi
\end{equation}
The perturbative $\beta$ function depends only on $m-2n$ to all 
orders~:
\begin{equation}
  \beta(g_{\sigma})=(m-2n-2)g_{\sigma}^{2}+O(g_{\sigma}^{3})
\end{equation}

The model for $g_{\sigma}$ positive thus flows to strong coupling for $m-2n>2$. 
Like in the ordinary sigma models case, the symmetry is restored at 
large length scales, and the field theory is massive. For $m-2n<2$ 
meanwhile, the model flows to weak coupling, and 
the symmetry is spontaneously  broken. One expects this scenario to 
work for $g_{\sigma}$ small enough, and the Goldstone phase to be 
separated from a non perturbative strong coupling phase 
by a critical point. 

\begin{figure}
\begin{center}
\includegraphics[scale=0.35]{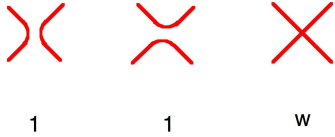}
%  \leavevmode
% \epsfysize=40mm{\epsffile{tensors.ps}}
\end{center}
\protect\caption{Interaction vertices in the $OSP(m|2n)$ symmetric
  sigma model on the square lattice.}
\label{fig:tensors}
\end{figure}

It is easy to suggest lattice models whose long distance physics is 
described by the supersphere sigma models. It was shown in \cite{JRS} 
that simply taking a square lattice with the fundamental 
representation of the $OSP(m|2n)$ on each link and Heisenberg
coupling at every vertex 
would suffice. More generallly, it is convenient to  represent the 
link states by lines carrying a label $a=1,\ldots,m$ or 
$\alpha=1,\ldots,2n$ and express the interactions in terms of the 
three invariant tensors of the algebra, corresponding to the three 
diagrams in figure~\ref{fig:tensors}.

A trivial rescaling of the partition function and  isotropy of the model 
allows one to set the first two weights equal to $1$. The remaining 
weight is a free bare parameter. It was found in \cite{JRS} that 
for {\sl any} $w>0$ and $m-2n<2$ the lattice model is indeed described at 
large distance by the UV limit of the sigma model, ie a set of $m-1$ 
free uncompactified bosons and $n$ pairs of symplectic fermions, with 
a central charge $c=m-2n-1$. The value $w=0$ is critical, and was 
argued in \cite{JRS,RS} to  correspond to the 
critical point of the sigma model. Note that for $m-2n$ fixed, there 
exists a particular value of the crossing weight $w$ where the model 
is integrable, and coincides with the one in \cite{nienhuis98}.

Of course, the geometrical representation of the invariant tensors 
allows for a full geometrical formulation of the model, that can then 
be extended to values of $m,n$ not integer. The model so obtained is 
made of loops covering every link of the lattice, and possibly
self-intersecting once on the vertices, with a fugacity $m-2n$, and a 
weight $w$ per crossing. The continuum limit was found to be a naive 
extrapolation from the results at $m,n$ integer. Note that this model 
differs from the usual formulation of the $Q$-state critical Potts 
model with $m-2n\equiv \sqrt{Q}$ only in that crossings are allowed. 
This however changes the universality class deeply. For instance, 
all the  $L$-leg operators which have non trivial, $Q$-dependent scaling 
dimensions in the Potts model for $-2\leq \sqrt{Q}\leq 2$, now have 
logarithmic correlators with vanishing effective dimension. This is 
because the symmetry being spontaneously broken the fundamental field 
has non vanishing expectation value, and thus the fields $\phi_{a}$ do 
exist in the conformal field theory, unlike in the case of a compact 
boson.

[Paths that can self intersect at a vertex but not at a bond are called 
{\sl trails} in the literature, and ``bond self avoiding models'' by 
contrast with ``site self avoiding''. The case $m-2n=0$ we consider 
below
would correspond to ``fully packed trails''. People have considered 
ordinary (dilute) trails which are in the same universality class as 
dilute SAW and trails at the theta point. Those have been shown to be 
in the same universality class as ``growing self avoiding trails'', 
themselves a particular case of the trajectory of a particle moving on 
a lattice with random distribution of scattering rotators. 
See \cite{Owczarek,Cao}. In the context of such trajectories, 
it has been remarked already that when one dilutes the set of random 
rotators from maximum concentration $C_{L}=C_{R}=1/2$ (i.e., add 
intersections), the exponents  
change.]

\subsection{Coupling constant and physical exponents of the $OSP(2|2)$ model}

We now  specialize to the  case of interest here, the 
$OSP(2|2)$ model. The UV limit is easy to understand.  We
can parametrize the supersphere
\begin{equation}
  (\phi_1)^2 + (\phi_2)^2 + 2\psi_{1}\psi_{2}=1
\end{equation}
by setting 
\begin{eqnarray} 
  \phi_{1} &=& \cos \phi \ (1-\psi_{1}\psi_{2}) \notag \\
  \phi_{2} &=& \sin \phi \ (1-\psi_{1}\psi_{2}),\qquad \phi \equiv \phi+2\pi
\end{eqnarray}
The action then reads
\begin{equation}
  S = \frac{1}{2g_{\sigma}} \int 
  d^{2}x \left[ (\partial_{\mu}\phi)^{2}(1-2\psi_{1}\psi_{2})
    +2\partial_{\mu}\psi_{1}\partial_{\mu}\psi_{2}
    -4\psi_{1}\psi_{2}\partial_{\mu}\psi_{1}\partial_{\mu}\psi_{2} \right]
\end{equation}
The coupling $g_{\sigma}>0$ flows to zero at large distances. On the other 
hand, we can absorb it by rescaling all fields so the action reads
\begin{equation}
  S = \frac{1}{2} \int d^{2}x 
  \left[(\partial_{\mu}\phi)^{2}(1-2g_{\sigma}\psi_{1}\psi_{2})+
    2\partial_{\mu}\psi_{1} \partial_{\mu}\psi_{2}
    +4g_{\sigma}\psi_{1}\psi_{2}\partial_{\mu}\psi_{1}\partial_{\mu}\psi_{2}\right]
\end{equation}
where now $\phi$ has a different radius, 
$\phi\equiv \phi+\frac{2\pi}{\sqrt{g_{\sigma}}}$. 
We see that as $g_{\sigma}\rightarrow 0$ all interaction 
terms disappear and we get a free boson $\phi$ together with a pair 
of free symplectic fermions $\psi_{1,2}$. Moreover the radius of 
compactification goes to infinity in that limit, so the 
boson $\phi$ appears as non compact. 
  
This holds in the true large distance limit. At intermediate 
scales, we can use the RG equation for the coupling \cite{AAR}
\begin{equation}
  \frac{dg_{\sigma}}{d\log l}=\frac{m-2n-2}{2\pi} g_{\sigma}^{2}
  = - \frac{1}{\pi} g_{\sigma}^{2}
\end{equation}
Writing more generally 
\begin{equation}
  \frac{dg_{\sigma}}{d\log l}= - \alpha g_{\sigma}^{2}
\end{equation}
we see that $g_{\sigma}$ approaches its vanishing large distance value as 
\begin{equation}
\frac{1}{g_{\sigma}} = \frac{1}{g_{\sigma}^{0}} + \alpha \log(l/l_0)
\simeq \alpha \log(l/l_0)
%  g_{\sigma}={g_{\sigma}^{0}\over g_{\sigma}^{0}+c\log l}\approx {1\over c\log l}
\end{equation}
Here, $l$ is a characteristic dimensionless scale ratio, roughly of 
the order of the ratio of the scale at which one is observing the 
physics to the lattice cut-off. On the cylinder, $l$ can be 
identified with the width in lattice units, $l=N$.

In the limit of large $l$, we can estimate more precisely the 
contribution to the spectrum coming from the boson $\phi$. 
Recall that for a free bosonic theory 
where the action is normalized as $S=\frac{1}{8\pi}\int 
(\partial_{\mu}X)^{2}$  and the field compactified as $X\equiv 
X+2\pi R$, the 
spectrum of dimensions \cite{DFMS} is
\begin{equation} \label{eq:gap-free-boson}
  \Delta+\bar{\Delta}=\frac{e^{2}}{R^{2}}+\frac{m^2 R^2}{4}
\end{equation}
Matching the normalization gives $R^{2}=4 \pi / g_{\sigma}$ in 
our case, and thus we expect the scaled gaps coming from the bosonic 
degrees of freedom to read, at large distances~:
\begin{equation}
  \Delta+\bar{\Delta}=\frac{e^{2}}{4\pi \alpha \log (l/l_0)}+m^{2}\pi \alpha \log (l/l_0)
\end{equation}
In the limit $l\rightarrow\infty$ the dimensions become 
degenerate and the spectrum can be considered as a continuum starting 
above $\Delta+\bar{\Delta}=0$. To emphasize the latter point, 
consider the contribution to the partition function coming from the 
$\phi$ degrees of freedom. The system is defined on a torus of periods
$l$ and $l'$, with $l'/l=\tau$. Denote $q = \exp(2i \pi \tau)$.
\begin{eqnarray}
  Z_{\phi}&=& \frac{1}{\eta \bar{\eta}}
  \sum_{e,m} q^{(e/R+mR/2)^{2}/2} \ \bar{q}^{(e/R-mR/2)^{2}/2} \notag\\
  &=& 
  \frac {R}{\sqrt{2}} \frac{1}{\sqrt{\hbox{Im }\tau} \ \eta \bar{\eta}}
  \sum_{m,m'} \exp \left(-\frac{\pi R^{2}|m\tau-m'|^{2}}{2 \ \hbox{Im }\tau}\right) \notag\\
  &\begin{array}{c}
    \approx\\
    R \rightarrow \infty \end{array} & \frac{R}{\sqrt{2}}
  \frac{1}{\sqrt{\hbox{Im }\tau} \ \eta\bar{\eta}}\label{zphi}
\end{eqnarray}
where $\eta=q^{1/24}\prod_{n=1}^{\infty}(1-q^{n})=q^{1/24}P(q)$. 
Observe now that  one can write~:
\begin{equation}
  \frac{1}{\sqrt{\hbox{Im }\tau} \ \eta\bar{\eta}}= 4\int_{0}^{\infty} 
  \frac{ q^{s^{2}}\bar{q}^{s^{2}}}{\eta\bar{\eta}} \ ds
\end{equation}
which can be interpreted as an integral over a continuum of critical 
exponents $\Delta=\bar{\Delta}=s^{2}$. In the partition function 
(\ref{zphi}), $R$  plays the role of the density of levels, 
and is proportional to the (diverging) size of the target space. 
 
Going back to the specific case of the $OSP(2|2)/OSP(1|2)$ model, we 
set $\alpha=1 / \pi$. We get the radius $R^{2}=4\log (l/l_{0})$, 
and the contribution of the free boson $\phi$ to the spectrum is~:
\begin{equation}
  \Delta+\bar{\Delta}=\frac{(e_2)^{2}}{4\log (l/l_0)}+(m_2)^{2}\log (l/l_0)
\end{equation}
with $e_2,m_2$ arbitrary integers. Suppose now we consider the sigma 
model with a combination of periodic and 
antiperiodic boundary conditions for the symplectic fermions. These 
get untwisted, and add to $Z_{\phi}$ the discrete spectrum of a free boson 
at the free fermion point, which is given by 
equation~\eqref{eq:gap-free-boson} with a radius
$R^{2}=1$. The quantum numbers associated to the fermions will be denoted 
$e_{1},m_{1}$. Thus we expect~:
\begin{equation}
  \Delta + \bar{\Delta} = {(e_1)^2} + \frac{(m_1)^2}{4}
  + \frac{(e_2)^2}{4\log (l/l_0)} + (m_2)^2 \log (l/l_0) \label{ospexp}
\end{equation}
We can now compare with formula~\eqref{eq:bethe-spectrum}. It is not 
entirely clear how constrained the quantum numbers might be in the 
lattice realization we are considering. But observe that finite scaled 
gaps occur for $m_+ = m_- = m$ and converge to 
\begin{equation}
  x=\frac{m^2}{g} + \frac{1}{4} g (n_+ + n_-)^{2} 
  + \frac{1}{4} K(\gamma, N) (n_+ - n_-)^{2} 
\end{equation}
When $\gamma$ tends to $\pi / 2$,
equations~\eqref{eq:g} and \eqref{eq:K-scaling-pi-2} give the coupling
constants~:
\begin{equation}
g=1 \ , \qquad K(\pi / 2, N) \sim \frac{A'}{\log N}
\end{equation}
The numerical results are compatible with $A'=1$.
This suggests the identifications~:
\begin{eqnarray}
e_1 &=& \frac{1}{2} (m_+ + m_-) = m \notag \\
m_1 &=& n_+ + n_- \notag 
\end{eqnarray}
\begin{eqnarray}
e_2 &=& n_+ - n_- \notag \\
m_2 &=& \frac{1}{2} (m_+ - m_-) = 0 \notag 
\end{eqnarray}
Note that the quantum numbers are (weakly) correlated :
the integers $m_1, e_2$ are such that $(m_1+e_2)$ is even.

\subsection{Interpretation of the staggered six-vertex model for
  $\gamma<\pi/2$}

In preparation for the subsequent discussion, we will 
denote the effective (square) radius of the non compact 
boson  by $(R_{2})^{2}$ (equal to $4\log (l/l_0)$ for $\gamma=\pi/2$)
and the radius of the compact  boson by  
$(R_1)^2$ (equal to $1$ for $\gamma=\pi/2$). Away from
$\gamma=\pi/2$, it would be  reasonable to try and 
interpret our results in terms of a deformation of the supersphere 
sigma model. Several scenarios are \textit{a priori} possible. From the 
numerical results, the compact direction clearly has a modified 
radius which becomes~:
\begin{equation}
  (R_{1})^{2} = g = \frac{2 \gamma}{\pi}
\end{equation}
As for the non compact direction, a first 
possibility would be to consider a constant $\alpha(\gamma)$ in the 
foregoing discussion. Although it provides reasonable results, much 
better fits are obtained with a radius going like $\log (l/l_0)$ for 
$\gamma \neq \pi/2$; specifically, we find~:
\begin{equation}
  (R_{2})^{2}=\frac{\pi -2\gamma}{5\gamma} [\log (l/l_0)]^{2}
\end{equation}
Note that this becomes ill-defined as $\gamma$ tends to $\pi/2$,
where there is crossover to a behaviour linear in $\log (l/l_0)$.
 
The most naive interpretation of the corresponding target space would 
be a torus with one period diverging like $\log (l/l_0)$ in the scaling limit. 
This is reminiscent of results on the sausage model \cite{FOZ} 
which is a deformation of the usual sphere sigma model. There 
however, the theory instead of being massless in the IR is massless 
in the UV, while  the  target space in that limit is asymptotically a cigar. 
The dependence of the cigar dimensions on the RG scale and the 
anisotropy parameter are however reminiscent of ours; in particular 
the ``long dimension'' goes as the logarithm of the RG coordinate in 
both cases.
 
Another, more suggestive  interpretation, can be obtained if we 
consider the partition function of our model on the torus. 
Set  $\gamma=\pi / t$ so $(R_{2})^{2}=\frac{t-2}{5}[\log (l/l_0)]^{2}$. Using the 
continuum representation (\ref{zphi}) and  
reabsorbing  the $t-2$  prefactor that comes from the dependence 
of $R_{2}$ upon $t$ into the 
continuously varying exponents we obtain the partition function~:
\begin{equation}
  Z \propto \frac{\log(l/l_{0})}{(\eta \bar{\eta})^2}
\sum_{e,m=-\infty}^{\infty} \int_{0}^{\infty} ds 
  \ q^{{s^{2}\over t-2}+{(m-te)^{2}\over 4t}}
  \ \bar{q}^{{s^{2}\over t-2}+{(m+te)^{2}\over 4t}} 
\end{equation}
where the proportionnality constant is a presumably non universal 
quantity, independent of $t$, equal appproximately to $1/\sqrt{10}$.
Observe now that ${2(t+1)\over t-2}-{24\over 4(t-2)}=2$ so the 
conformal weights can  be written as well, with respect to 
$c={2(t+1)\over t-2}$, as~:
\begin{equation}
  h={s^{2}+{1\over 4}\over t-2}+{(m\mp te)^{2}\over 4t}
\end{equation}
This  coincides with the contribution of the continuous representations 
to the spectrum of the Euclidian black hole CFT $SL(2,R)/U(1)$ coset model
\cite{Dijkgraaf}.

It might be useful here to give some quick reminders for the 
spectrum of this model. The central charge for level $k$ is~:
\begin{equation}
  c_{\rm BH}={2(k+1)\over k-2}
\end{equation}
Normalizable operators come in two kinds. There is the ones (delta 
functions normalizable) associated 
with the principal continuous series $j=-{1\over 2}+is$ with conformal 
weight~:
\begin{equation}
  h=-{j(j+1)\over k-2}+{(n\pm kw)^{2}\over 4k}={s^{2}+{1\over 4}\over 
    k-2}+{(n\pm kw)^{2}\over 4k}\label{continuous}
\end{equation}
and the ones coming from the discrete series. They have 
$j \in \left[{1-k\over 2},-{1\over 2}\right]$ together with  rules relating 
allowed values to $w,n$. The net result is that these fields all have 
dimensions larger than the bottom of (\ref{continuous}). Note that 
the identity field $h=0$ is not among the normalizable states, which 
is consistent with the fact that we do not observe (after the obvious 
identification $k\equiv t$)  $c_{\rm BH}$ in the 
lattice model but $c_{\rm BH}-24 {1\over 4(k-2)}=2$. 

Of course the partition function of the Euclidian black hole theory 
should naively be infinite since it involves infinite dimensional 
representations of $SL(2)$. The introduction of a Liouville wall at 
finite distance in the target space \cite{Hanany} gives a density of 
states $\rho(s)\propto \log \epsilon$ to leading order, which agrees 
with our results provided we identify $\epsilon\equiv l/l_{0}$, and 
thus  the size of 
the system (it would certainly be interesting to investigate the 
subleading behaviour, which depends on $s$, in the lattice data).

Finally, we note that in \cite{Hanany}, the sum (4.17)  
implies some combinatorial contraints on the descendents, which we 
have not studied in the lattice model. 
 
Note that further twisting and reduction of the vertex model gives 
rise to parafermionic theories, which are themselves cosets 
$SU(2)/U(1)$. It is not clear how this might be related to the 
identification of the untwisted vertex model with $SL(2,R)/U(1)$. 
 
\section{Geometrical interpretation of the critical exponents}
\label{sec:geometrical}

For the $OSP(m|2n)$ models in their Goldstone phases, the continuous 
spectrum of critical exponents has its origin (like in the case of a 
pure 
non compact boson) in the existence of infinitely many operators 
with vanishing scaling dimension (the powers of $\phi$ in the case of 
the non compact boson), which in itself is a consequence of the 
spontaneously broken symmetry \cite{JRS}. An obvious question is 
whether a similar interpretation exists for our model.

Some insight can be gained by considering the limit 
$\gamma \to \pi/2$ which is related to the $OSP(2|2)$ model. In 
the latter, exponents of the order parameter are related with 
geometrical correlations of the degrees of freedom carrying lines, and 
therefore it is tempting to ask whether this might hold away from the 
limit point as well. Before discussing this, it is important to 
notice that the staggered six-vertex model at $\gamma=\pi/2$
(and the equivalent 38-vertex model) with periodic 
boundary conditions   correspond, strictly speaking,  to 
the  model of \cite{JRS,JSarboreal} where 
the $OSP$ symmetry is {\sl  broken} by the boundary. 
This is the reason why the spectrum of conformal weights contains 
a compact boson with $(R_1)^2=1$, and thus non trivial, finite 
exponents. Within the geometrical interpretation to be discussed 
below, this will correspond to the existence of some correlators 
having non trivial weights, while others do behave as in \cite{JRS}.
%thus should not be a surprise 

Let us now be more specific. 
As shown in section~\ref{sec:R-matrix}, summing on $2 \times 2$
vertex blocks and choosing the right basis, the staggered six-vertex 
model is equivalent to the 38-vertex model. Any lattice configuration
within this model is completely covered by polygons of three different
types~: oriented lines, bare thin lines and bare thick lines. The 
particular point $\gamma=\pi/2$, as well as our discussion on 
Goldstone phases, suggests to consider 
geometrical correlations associated with these lines.

\subsection{Correlation functions with no thin or thick line}

\begin{figure}
\begin{center}
\includegraphics[scale=0.5]{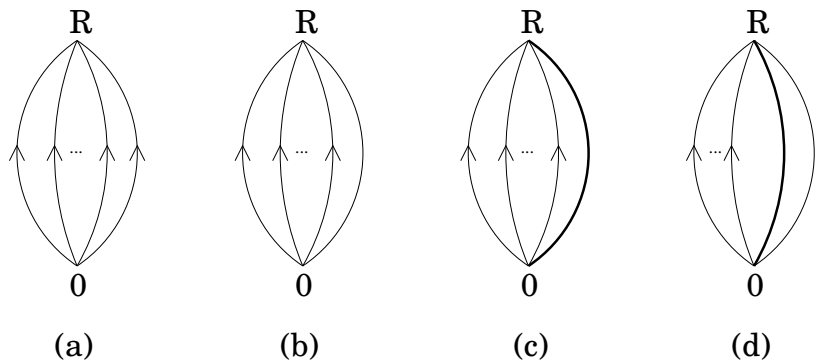}
\caption{Some ``watermelon'' correlation functions. (a): no bare line,
  (b)-(c): one bare line, (d): one thin line and one thick line.}
\label{fig:watermelon}
\end{center} 
\end{figure}

Consider first the watermelon correlation function (as illustrated on
the figure \ref{fig:watermelon})
consisting of a forced  even number $2r$ of positively-oriented lines (that 
is, we force $2r$ such lines to propagate through the system, on top
of fluctuating numbers of positively and negatively oriented lies in
equal number and bare lines).
The critical exponent of this correlation function
is given by the lowest-energy state in the sector with 
fixed total spin $S=2r$ and $C=1$.
According to equation~\eqref{eq:bethe-spectrum}, this state is 
$\varphi_{(r,r),(0,0)}$.
According to equation~\eqref{eq:action-C-2}, this is an eigenstate of
the operator $C$, with eigenvalue one.

\subsection{Correlation functions with one thin or thick line}

Consider the correlation function consisting of a forced  
odd number $2r-1$ of positively-oriented lines, along
with one bare (thin or thick) line. The critical exponent 
of the correlation function with a thin (resp. thick) line
is given by the lowest-energy 
state in the sector with fixed total spin $S=2r-1$ and $C=1$
(resp. $C=-1$).
Both states $\varphi_{(r, r-1), (0,0)}^{\pm}$ have the physical 
exponent $x_{(r, r-1), (0,0)}$. Thus, the correlation functions
with one thin or thick line have the same exponent 
$x_{(r, r-1), (0,0)}$.

\subsection{Correlation functions with one thin line and one thick line}

Consider the correlation function consisting of a forced 
even number $2r$ of positively-oriented lines, along
with one thin line and one thick line. The critical exponent 
of this correlation function is given by the lowest-energy 
state in the sector with fixed total spin $S=2r$ and $C= -1$.
This state is $\varphi_{(r+1, r-1), (0,0)}^{-}$, with physical
exponent $x_{(r+1, r-1), (0,0)}$.

In the particular case $r=0$, this exponent becomes~:
\begin{equation}
x_{(1,-1), (0,0)} = K(\gamma, N)
\end{equation}
This exponent vanishes in the thermodynamic limit.
So the two-leg correlation function consisting of one 
thin line and one thick line belongs to the continuous
subspectrum associated to the ground state.

\subsection{Interpretation}

The physical interpretation we suggest  for the continuum limit of 
the 38-vertex model is thus the following. The 
arrow degrees of freedom can be treated like the usual 
domain boundaries for a RSOS model which renormalizes, at large 
distances, to a compactified free bosonic field with radius $R_{1}$. A correlation 
function involving a certain number $S$ of positively oriented lines 
corresponds for this bosonic field to a magnetic or vortex operator,
whose physical exponent
is given by $x= g S^2 / 4$.

Meanwhile, correlators involving in addition thin and thick lines as
well have, in the thermodynamic limit, exponents entirely determined
from the contribution of the arrow degrees of freedom.%
\footnote{A subtle remark is in order here. Recall the 
 conserved quantities of the 38-vertex
model : the total spin and the value of operator $C$. In
geometrical terms, this means that the \emph{total} arrow flow
is conserved, whereas only the \emph{parity} of the number of thick
lines is conserved. To define a correlation function with more than
one thick lines within the 38-vertex partition function, it is
necessary to define the transfer matrix on a non-local Hilbert space
(so that the transfer matrix keeps track of the connectivity of 
the thick lines). In this framework, three interpretations are
possible for the vertex $a_1^{(8)}$. The behaviour of the correlation
functions is likely to depend on the ``splitting'' of this vertex into
the three possible connectivities, though we believe that the behaviour
is universal, in agreement with this conclusion. 
We did not enter into these technicalities here, and simply 
described the correlation functions which are not affected
by this problem.}
Thin and thick lines correspond thus to operators with vanishing
exponents and (presumably) logarithmic correlators, and behave
similarly to the crossing lines in the models of \cite{JRS}.

% [It is  intriguing to wonder how close these lines are  
% to Brownian curves  Recall that the 
%two point function of a non compact boson can be interpreted as a 
%sum over Brownian trajectories connecting the two points, with a 
%fixed fugacity $1/\hbox{ coordination number}$]. 

\section{Conclusion, open problems}

First and foremost, we believe this work puts on firm ground the 
existence of a continuous spectrum of critical exponents in a model 
with a finite number of lattice degrees of freedom per site (link), 
justifying  fully the conclusions of \cite{JRS,EFS} .

It is truly remarkable that a proper staggering of the simple six-vertex model 
could give rise to such  interesting behaviour: obvious directions 
for future work are plenty, and include an 
analytic derivation of the coupling constant $K$, a better 
understanding of the relationship with $SL(2,R)/U(1)$, and of the 
effect of the various twist and truncations necessary to produce the 
Potts and RSOS versions of this model. 

It should also be possible to generalize the problem to some higher 
spin version. Some comments on Bethe equations are in order here. 
Recall that the usual source term for these equations reads, in the 
case of spin $1/2$~:
\begin{equation}
\left({\sinh{1\over 2}(\alpha-i\gamma)\over \sinh{1\over 
2}(\alpha+i\gamma)}\right)^{N}
\end{equation}
One might think of changing it through real heterogeneities $\pm 
\Lambda$, leading to equations which have been used a great deal in the 
study of massive deformations \cite{Reshetikhin}~:
\begin{equation}
\left({\sinh{1\over 2}(\alpha-\Lambda-i\gamma)\over 
\sinh{1\over 2}(\alpha-\Lambda+i\gamma)}\right)^{N/2}
\left({\sinh{1\over 2}(\alpha+\Lambda-i\gamma)\over 
\sinh{1\over 2}(\alpha+\Lambda+i\gamma)}\right)^{N/2}
\end{equation}
One could also think of changing it through imaginary 
heterogeneities. The simplest case would correspond to adding ``string 
heterogeneities'', i.e., for the simplest case of the two string, 
formally $\Lambda=i\gamma$. Clearly however half the modified source 
terms cancel out, leaving~:
\begin{equation}
\left({\sinh{1\over 2}(\alpha-2i\gamma)\over \sinh{1\over 
2}(\alpha+2i\gamma)}\right)^{N}
\end{equation}
which is nothing but the source term for a spin one chain. In 
general, heterogeneities of the type p-string will lead to the 
equations for the spin $p/2$ chain.
The other natural possibility would be to add heterogeneities of the 
anti-string type, ie $\Lambda=i\pi$. Up to a shift $\alpha\rightarrow 
\alpha+i\pi$, this produces however the initial equations, and is 
well known to simply change the Hamiltonian from ferromagnetic to 
antiferromagnetic.
The possibility we encountered in this paper consists in adding 
heterogeneitiesright in the middle of the ``physical strip'', 
at  $\Lambda=i{\pi/2}$. Note that the higher spin 
generalization is obvious by changing $i\gamma$ to $2si\gamma$ 
everywhere. 

It is also fascinating that the model should interpolate between 
$OSP(2/2)$ and  $SO(4)$ symmetries when $\gamma$ goes from $\pi/2$ to 
zero. This suggests the existence of a possible quantum group 
symmetry all along the line, which we have unfortunately not yet been
able to identify.

\end{document}